\documentclass[%
reprint,
superscriptaddress,
 amsmath,amssymb,
 aps,
prd,
]{revtex4-2}

\usepackage{graphicx}
\usepackage{dcolumn}
\usepackage{bm}

\begin{document}


\title{Observing gravitational wave polarizations with LISA-TAIJI network}


\author{Gang Wang}
\email[Gang Wang: ]{gwang@shao.ac.cn, gwanggw@gmail.com}
\affiliation{Shanghai Astronomical Observatory, Chinese Academy of Sciences, Shanghai 200030, China}

\author{Wen-Biao Han}
\email[Wen-Biao Han: ]{Corresponding author:wbhan@shao.ac.cn}
\affiliation{Shanghai Astronomical Observatory, Chinese Academy of Sciences, Shanghai 200030, China}
\affiliation{School of Astronomy and Space Science, University of Chinese Academy of Sciences, Beijing 100049, China}

\date{\today}

\begin{abstract}

Two polarization modes of gravitational wave are derived from the general relativity which are plus and cross modes.
However, the alternative theories of gravity can yield the gravitational wave with up to six polarizations. Searching for the polarizations beyond plus and cross is an important test of general relativity. In principle, one space-borne detector, like LISA, could measure the gravitational wave polarizations from a long time observation with its orbital motion. With the comparable sensitivities, the joint LISA and TAIJI missions will improve the observations on the polarization predictions of theories beyond general relativity. In this work, a class of parameterized post-Einsteinian waveform is employed to describe the alternative polarizations, and six parameterized post-Einsteinian parameters quantifying from general relativity waveform are examined by using the LISA-TAIJI network. Our results show that the measurements on amplitudes of alternative polarizations from joint LISA-TAIJI observation could be improved by more than 10 times compared to LISA single mission in an optimal scenario.

\end{abstract}


\maketitle


\section{Introduction}

From O1 to O3a of the Advanced LIGO and Virgo runs, About 50 gravitational wave (GW) events have been reported \cite{GWTC-1PhysRevX.9.031040,gwtc-2}. The mergers of binary compact objects offer the unique chance to test general relativity (GR) in the extra strong and dynamical gravitational field \cite{ligotestgr2016PhRvL.116v1101A,ligotestgr1,ligotestgr2}. During these tests, the polarization of GWs is an important issue. The general relativity (GR) predicts only two tensor polarizations: plus (+) and cross ($\times$) modes. However, the metric theories of gravity may yield up to six polarizations which are two vector modes, two scalar modes, 
and the two transverse-traceless polarization modes in GR. For examples, scalar-tensor theories like as Brans-Dicke theory predict an extra scalar polarization (breathing, b) mode\cite{brans-dicke1961PhRv,Will2014LRR}; Vector-tensor theories can excite vector modes (x, y modes) \cite{will1993book}; Einstein-Aether theory \cite{Einstein-aether2004} predicts the existence of five polarization modes; tensor-vector-scalar theories such as TeVeS\cite{teves2004PhRvD}, bimetric\cite{rosenPhysRevD.3.2317,ROSEN1974455} and stratified theories such as Lightman-Lee theory\cite{lightmanleePhysRevD.8.3293}, predict the existence of all six polarization modes (+, $\times$, x, y, b, L), where L mode means longitudinal and is another scalar polarization mode.

In general, for the transient GW signals, at least three detectors are required to constrain additional modes\cite{yunes2012PhRvD..86b2004C}. Four detectors are necessary to constrain the vector modes, and in order to fully disentangle the polarization content of a transient signal, at least five detectors are needed to break all degeneracies\cite{yunes2012PhRvD..86b2004C,ligotestgr2}. After the Advanced Virgo joined the GW observation network, the tensor polarizations have been tested if they are preferred over other modes with GW170814 and GW170817 \cite{GW170814PhysRevLett.119.141101,Abbott:2018lct}. The KAGRA in Japan has begun operating and will improve the measurement of polarizations in the near future \cite{takedaPhysRevD.98.022008,takedaPhysRevD.100.042001,hagiharaPhysRevD.100.064010,takeda2020arXiv201014538T,kagra2020arXiv200802921K}. 

The space-borne detectors including LISA \cite{2017arXiv170200786A}, TAIJI \cite{Hu:2017mde}, and TianQin \cite{Luo:2015ght} are planed to been launched around the 2030s targeting to detect the GW in the low frequency band. As a benefit of the periodical motions orbiting the Sun/Earth, the detectors can observe from (long-lasting) GW signals at different positions and orientations. And then one single mission, like LISA, could measure the polarizations independently \cite{Nishizawa2010PhRvD,Isi2015PhRvD}, especially for the sources at optimal position and inclination. However, for the massive black hole binaries (MBHBs), the duration of the signal is about a few weeks and may not be enough to constrain the polarizations. The joint observation from LISA-TAIJI network may improve resolution of the measurements. With the comparable sensitivities of two missions, there will be lots of merits by LISA-TAIJI joint observations. \citet{Ruan:2020smc} and \citet{Wang:2020a} demonstrated a significant improvement in sky localization capacities by LISA-TAIJI network. \citet{Omiya:2020fvw} and \citet{Orlando:2020oko} calculated the overlap reduction functions of the two missions and evaluated the impacts of the joint observations on the stochastic GW observation. \citet{Liu:2020mab} estimated the constraint on polarizations from single TAIJI observations.

In this paper, following our previous work in \cite{Wang:2020a}, by using the LISA-TAIJI network, we evaluate the capacity of observation for the polarization predictions beyond general relativity.  A set of post-parameterized Einsteinian (ppE) waveform is employed to represent the GW signals of six potential polarization, and six ppE parameters are used to quantify the deviations of GW from GR. The Fisher information matrix algorithm is utilized to determine measurements on the six parameters from two MBHB sources. The results show that the measurements on amplitudes of alternative polarizations from joint LISA-TAIJI observation could be improved by more than tenfold compared to LISA single mission in an optimal scenario.

This paper is organized as follows. 
In Sec. \ref{sec:ppE_formulation}, we introduce the model independent waveforms with GW and alternative polarization modes. 
In Sec. \ref{sec:response}, we specify the responding functions of the time-delay interferometry (TDI) to the polarizations, and evaluate the average sensitivities for the different polarization modes. 
The Fisher information matrix method utilized and the determinations on the ppE parameters are presented in Sec. \ref{sec:SMBH_results}.
We recapitulate our conclusions in Sec. \ref{sec:conclusions}.
(We set $G=c=1$ in this work).

\section{Parameterized post-Einstein waveforms with all polarizations} \label{sec:ppE_formulation}

The GWs derived from GR have only two polarization modes $h_+$ and $h_\times$, the time-domain waveforms from a binary system with quadrupole approximation are
\begin{align}
h_{+} &=-\frac{2 \mu M}{D r} \cos 2 \Phi\left(1+\cos ^{2} \iota\right), \\
h_{\times} &=-\frac{4 \mu M}{D r} \sin 2 \Phi \cos \iota,
\end{align}
where $M$ the total mass, $\mu$ is the reduced mass $\frac{m_1 m_2}{m_1 + m_2}$,  $r$ the separation of two bodies, $\Phi$ the orbital phase, $D$ is the luminosity distance, and$\iota$ inclination angle of the source with respective to the light-of-sight. The responded signal in a detector will to the GW be
\begin{align}
    h_{\rm GR}(t)=F_{+} h_{+}+F_{\times} h_{\times}, \label{f2h}
\end{align}
where $F_+$ and $F_\times$ are antenna pattern functions of the detector to the two polarizations. On the other sides, the frequency evolution of a binary under PN approximation is a classical solved problem \cite{peters1963PhRv..131..435P,peters1964PhRv..136.1224P}, and the GW from GR in frequency-domain could be approximated as \cite{gwbookI}
\begin{align}
\tilde{h}_{\mathrm{GR}}(f)=\left(\frac{5 \pi}{96}\right)^{1 / 2} A_{\mathrm{GR}} \frac{\mathcal{M}^{2}}{D}(\pi \mathcal{M} f)^{-7 / 6} e^{-i \Psi_{\mathrm{GR}}},
\end{align}
where $\mathcal{M}$ is the chirp mass of the binary $(m_1 m_2)^{3/5} / (m_1 + m_2)^{1/5}$, $A_{\rm GR}$ is the responding amplitude of polarization modes ($+,~\times$) from a detector
\begin{align}
    A_{\rm GR} = -F_+(1+\cos^2\iota)-2i F_\times \cos\iota.
\end{align}

In general, GW metric perturbations at a given space-time point can be expressed as 
\begin{align}
    h_{i j}(t, \hat{\Omega})=h_{a}(t) e_{i j}^{a}(\hat{\Omega}), \label{gwperturbation}
\end{align}
where $\hat{\Omega}$ is the sky direction of a GW source. In metric theories of gravity, there are up to six possible polarization modes because the polarization tensors $e_{i j}^{a}(\hat{\Omega})$ could have maximum six combinations which are defined as
\begin{align}
e_{a b}^{+}&=\hat{e}_{x} \otimes \hat{e}_{x}-\hat{e}_{y} \otimes \hat{e}_{y}, \\ 
e_{a b}^{\times}&=\hat{e}_{x} \otimes \hat{e}_{y}+\hat{e}_{y} \otimes \hat{e}_{\rm x},\\ 
e_{a b}^{\rm x}&=\hat{e}_{x} \otimes \hat{e}_{z}+\hat{e}_{z} \otimes \hat{e}_{x},\\ 
e_{a b}^{\rm y}&=\hat{e}_{y} \otimes \hat{e}_{z}+\hat{e}_{z} \otimes \hat{e}_{y},\\ 
e_{a b}^{\rm b}&=\hat{e}_{x} \otimes \hat{e}_{x}+\hat{e}_{y} \otimes \hat{e}_{y},\\ 
e_{a b}^{\rm L}&=\sqrt{2} \hat{e}_{z} \otimes \hat{e}_{z},
\end{align}
where the set of orthonormal unit vectors $\{\hat{e}_{x}, ~\hat{e}_{y},~\hat{e}_{z}\}$ is GW basis, i.e., $\hat{e}_{z} = -\hat{\Omega}$ is a unit vector in the direction of propagation of the GW and $\hat{e}_{z} = \hat{e}_{x} \times \hat{e}_{y}$. Then in Eq. (\ref{gwperturbation}), $a = +,~\times, ~\rm{x},~\rm{y},~\rm{b},~\rm{L}$ are the polarization indices and corresponding to the two tensor modes ($h_+, ~h_\times$),  two vector modes ($h_{\rm x}, ~h_{\rm y}$) and two scalar modes (breathing $h_{\rm b}$ and longitudinal $h_{\rm L}$). 

By assuming all six polarization modes exist, an observed GW signal in a detector can be written as 
\begin{align}
    h(t)=F_{+} h_{+}+F_{\times} h_{\times}+F_{\rm x} h_{\rm x}+F_{\rm y} h_{\rm y}+F_{\mathrm{b}} h_{\mathrm{b}}+F_{\mathrm{L}} h_{\mathrm{L}}. \label{f6h}
\end{align}
Similar to Eq. (\ref{f2h}), the $F_{\rm x}, ~F_{\rm y}, ~F_{\mathrm{b}}$, and $F_{\mathrm{L}}$ are the response functions of the detector to the extra polarization modes beyond the GR. However, the $h_{\rm x},~h_{\rm y},~ h_{\mathrm{b}}$ and $h_{\mathrm{L}}$ are the GW waveform for the corresponding polarizations, and they are derived in a different forms from various theories. Also not all the polarization modes appears in different theories. For instance, the Brans-Dicke theory (a scalar-tensor theory) only predicts one more mode $h_{\rm b}$. Vector-tensor theories usually predict the existence of preferred directions and the excitation of vector modes $h_\mathrm{x},~h_\mathrm{y}$ \cite{will1993book}. Einstein-Aether theories will allow 5 polarization modes \cite{Einstein-aether2004},  while the tensor-vector-scalar theories, bimetric and stratified theories allow all 6 polarization modes \cite{teves2004PhRvD,rosenPhysRevD.3.2317,ROSEN1974455,lightmanleePhysRevD.8.3293}. The above alternative gravity theories predict different formalism for a certain extra polarization. One can refer to Ref. \cite{yunes2012PhRvD..86b2004C} the time-domain waveforms with different coupling parameters for a few of these theories. For the testing of these polarization modes, it is not convenient to focus on just one special theory. Alternatively, a model-independent waveform model which can include a set of parameters corresponding to various theories should be more appropriate for tests of polarization.

Furthermore, for data analysis, frequency-domain waveforms are will be more convenient the time-domain ones. And the waveforms in frequency-domain have different formulas for different theories not only due to the time-domain waveforms but also due to varied radiation reactions. 
A general framework which can incorporate the possible alternative theories of gravity will be convenient to test the potential polarizations beyond the GR.

To test the GR in the post-Newtonian limits, the parameterized post-Newtonian (ppN) formalism was developed in the 1970s \cite{Thorne:1970wv,Will:1971zzb,Ni:1972wx}. The ppN formalism provided a good approach on tests of gravity theories in the solar system, binary pulsars, motion of objects around supermassive black hole, and etc \cite{Will2014LRR}. 
To test the alternative theories of gravity beyond the GR, by adopting the similar strategy like ppN, \citet{ppe2009PhRvD..80l2003Y} developed a parameterized post-Einsteinian (ppE) formalism to incorporate the alternative theories beyond GR. \cite{yunes2012PhRvD..86b2004C} extended a model-independent framework to including the complete polarization content. The ppE approach provides a approach for testing GR through the GW observations.

The standard ppE waveform in frequency domain can be generally expressed as
\begin{align}
    \tilde{h}(f) = \tilde{h}_{\rm GR}(f) [1+\alpha' u^{a'}] e^{i\beta' u^{b'}},
\end{align}
where ($\alpha', ~a'$) are ppE parameters on the amplitude modification, and ($\beta',~b'$) are parameters on phase correction, here we use the superscript $'$ to distinguish from the following parameters $\alpha,~\beta$, and $b$ which have coefficient differences defined in \cite{yunes2012PhRvD..86b2004C}. $u = \pi \mathcal{M}f$ when the dominant GW mode is considered. The waveform will return to the PN waveform described in GR $\tilde{h}_{\rm GR}(f)$ as when the ppE parameters go to zero.

When full six possible polarization modes are considered, by taking the waveform from harmonic $l =2$, a model-independent ppE framework from \cite{yunes2012PhRvD..86b2004C} are described as
\begin{equation}
\begin{aligned}
    \tilde{h}_{\mathrm{ppE}}(f) = & \tilde{h}_{\mathrm{GR}}\left(1+c \beta u_{2}^{b+5}\right) e^{2 i \beta u_{2}^{b}}+\left[\alpha_{\mathrm{b}} F_{\mathrm{b}} \sin ^{2} \iota \right. \\ 
   & \left.  +\alpha_{\mathrm{L}} F_{\mathrm{L}} \sin ^{2} \iota+\alpha_{\rm x} F_{\rm x} \sin 2 \iota+\alpha_{\rm y} F_{\rm y} \sin \iota\right] \\
   &\times \frac{\mathcal{M}^{2}}{D} u_{2}^{-7 / 2} e^{-i \Psi_{\mathrm{GR}}^{(2)}} e^{2 i \beta u_{2}^{b}}, \label{ppe6waveform}
\end{aligned}
\end{equation}
where $u_2 \equiv (\pi \mathcal{M}f)^{1/3}$ as defined in this case, $\beta$ and $b$ are the free ppE parameters, $c$ is a coefficient decided by $b$. $\alpha_{\rm b}, ~\alpha_{\rm L}, ~\alpha_{\rm x}$, and $\alpha_{\rm y}$ are the parameters related to the breathing, longitudinal and vector polarization x and y modes, respectively. $F_{\rm b, L, x, y}$ are the response functions of one GW detector to each corresponding polarization mode.  $c$ is defined as follow to incorporate the conservative and dissipative corrections as defined by Eq. (11) in the Erratum \cite{PhysRevD.95.129901} of \cite{yunes2012PhRvD..86b2004C},
\begin{align}
    c=-\frac{16}{15} \frac{b(3-b)\left(b^{2}+7 b+4\right)}{b^{2}+8 b+9}.
\end{align}
The relation between $b$ and $c$ means that the modification on the GW phase will definitely influences on the amplitude as analyzed in \citep{yunes2012PhRvD..86b2004C}.
The GW approximation in this work is based on the Eq. (\ref{ppe6waveform}), and the antenna pattern functions for polarization mode p, $F_\mathrm{p}$ for a LISA-like detectors will be specified in the next section.

\section{GW Response in TDI} \label{sec:response}

\subsection{The LISA and TAIJI orbital configuration}

\begin{figure}[htb]
\includegraphics[width=0.48\textwidth]{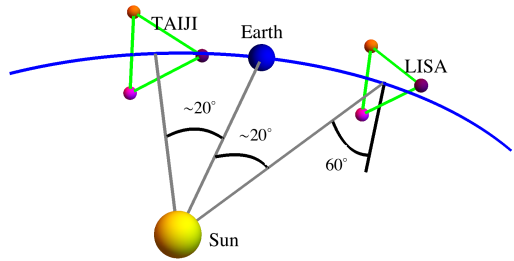}
\caption{\label{fig:LISA_TAIJI} The diagram of LISA and TAIJI mission orbital configurations. }
\end{figure}

The updated LISA mission proposed a $2.5 \times 10^6$ km arm length and trails the Earch by around $20^{\circ}$ \cite{2017arXiv170200786A}.
The formation plane of the three S/C would has 60$^{\circ}$ inclination angle with respect to the ecliptic plane as shown in Fig. \ref{fig:LISA_TAIJI}. The TAIJI missio proposed a LISA-like formation which $3 \times 10^6$ km arm length \cite{Hu:2017mde}. The triangle constellation will be in front of the Earth by around 20$^{\circ}$ as shown in Fig. \ref{fig:LISA_TAIJI}. 

With a separation of $\sim1 \times 10^8$ km, and the joint LISA-TAIJI observation from the long baseline will bring merits for GW detections. In our previous work \cite{Wang:2020a}, we evaluated the sky localization improvement of the joint observation on the suppermassive black hole binaries. By employing the numerical mission orbit in \cite{Wang:2017aqq,Wang:2020a}, we will explore the detectability of the joint network to the alternative GW polarization modes beyond the GR.

\subsection{Michelson and optimal TDI channels}

\begin{figure}[htb]
\includegraphics[width=0.23\textwidth]{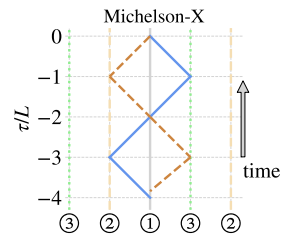}
\caption{\label{fig:X_A_E_T} The S/C layout-time delay diagrams for Michelson X channels as generated in \cite{Wang:2020cpq}. }
\end{figure}

The optimal channels of the first-generation Michelson TDI channel are employed to perform the detectability of the LISA/TAIJI mission. The Michelson X channel spacecraft (S/C) layout-time delay diagram is shown in Fig. \ref{fig:X_A_E_T} as generated in \cite{Wang:2020cpq}.
Following the diagram, the expression of measurements in the X channel will be \cite{Vallisneri:2012np}. 
\begin{equation} \label{eq:X_measurement}
\begin{aligned}
{\rm X} =& [ \mathcal{D}_{31} \mathcal{D}_{13} \mathcal{D}_{21} \eta_{12}  + \mathcal{D}_{31}  \mathcal{D}_{13} \eta_{21}  +  \mathcal{D}_{31} \eta_{13} +  \eta_{31}   ] \\
& - [ \eta_{21} + \mathcal{D}_{21} \eta_{12} +\mathcal{D}_{21} \mathcal{D}_{12} \eta_{31} + \mathcal{D}_{21}  \mathcal{D}_{12} \mathcal{D}_{31} \eta_{13} ] \\
\end{aligned}
\end{equation}
where $\mathcal{D}_{ij}$ is a time-delay operator, $ \mathcal{D}_{ij} \eta(t) = \eta(t - L_{ij} ) $.
The $\eta_{ji}$ are the combined observables from S/C$j$ to S/C$i$ which are defined as \citep{Otto:2012dk,Otto:2015,Tinto:2018kij}, and the specific expressions for this work are defined by Eq. (2) in \cite{Wang:2020cpq}.

A group of optimal TDI channels, (A, E, and T), can be generated from linear combinations of the three Michelson channels (X, Y, and Z) as following \cite{Prince:2002hp,Vallisneri:2007xa}, 
\begin{equation} \label{eq:optimalTDI}
 {\rm A} =  \frac{ {\rm Z} - {\rm X} }{\sqrt{2}} , \quad {\rm E} = \frac{ {\rm X} - 2 {\rm Y} + {\rm Z} }{\sqrt{6}} , \quad {\rm T} = \frac{ {\rm X} + {\rm Y} + {\rm Z} }{\sqrt{3}}.
\end{equation}
The Y and Z channels are obtained from cyclical permutation of the S/C indexes. The joint three optimal channels would represent the ultimate detectability of a LISA-like space mission. Therefore, the joint optimal channels are employed to study the capability of the LISA and TAIJI mission to the GW signals.

\subsection{Response formulation of TDI channel} \label{secsub:TDI_response}

The final GW response of a TDI channel is combined from the response in each single link.
The response to a GW $+$ and $\times$ polarizations in a single link Doppler measurement has been formulated in \cite{1975GReGr...6..439E,1987GReGr..19.1101W}, and specific formulas were described in \citet{Vallisneri:2007xa,Vallisneri:2012np}. \citet{Tinto:2010hz} developed the response functions for the alternative polarizations and evaluated the sensitivities. We employ the formulas as follows to investigate the response of TDI to the six polarizations.

For a GW source locating at ecliptic longitude $\lambda$ and latitude $\theta$ with respect to the solar-system barycentric coordinates, the GW propagation vector will be
\begin{equation} \label{eq:source_vec}
 \hat{k}  = -( \cos \lambda \cos \theta, \sin \lambda \cos \theta ,  \sin \theta ).
\end{equation} 
The $+$ or $\times$ polarization tensors of the GW signal, as well as the (potential) alternative polarization tensor, scalar breathing (b), scalar longitudinal (L), vector x and y, combining with the factors from inclination angle $\iota$ of the source are
\begin{widetext}
\begin{equation} \label{eq:polarizations-response}
\begin{aligned}
{\rm e}_{+} & \equiv \mathcal{O}_1 \cdot 
\begin{pmatrix}
1 & 0 & 0 \\
0 & -1 & 0 \\
0 & 0 & 0
\end{pmatrix}
\cdot \mathcal{O}^T_1 \times \frac{1+\cos^2 \iota}{2} ,
\qquad
{\rm e}_{\times}  \equiv \mathcal{O}_1 \cdot 
\begin{pmatrix}
0 & 1 & 0\\
1 & 0 & 0 \\
0 & 0 & 0
\end{pmatrix}
\cdot \mathcal{O}^T_1 \times i (- \cos \iota ), 
\\
{\rm e}_\mathrm{b} & \equiv \mathcal{O}_1 \cdot 
\begin{pmatrix}
1 & 0 & 0\\
0 & 1 & 0 \\
0 & 0 & 0
\end{pmatrix}
\cdot \mathcal{O}^T_1 \times \sin^2 \iota,  
\qquad \qquad \quad
{\rm e}_\mathrm{L}  \equiv \mathcal{O}_1 \cdot 
\begin{pmatrix}
0 & 0 & 0\\
0 & 0 & 0 \\
0 & 0 & 1
\end{pmatrix}
\cdot \mathcal{O}^T_1 \times \sin^2 \iota,
\\
{\rm e}_\mathrm{x} & \equiv \mathcal{O}_1 \cdot 
\begin{pmatrix}
0 & 0 & 1\\
0 & 0 & 0 \\
1 & 0 & 0
\end{pmatrix}
\cdot \mathcal{O}^T_1 \times \sin \iota \cos \iota, 
\qquad \quad \ 
{\rm e}_\mathrm{y}  \equiv \mathcal{O}_1 \cdot 
\begin{pmatrix}
0 & 0 & 0\\
0 & 0 & 1 \\
0 & 1 & 0
\end{pmatrix}
\cdot \mathcal{O}^T_1  \times i \sin \iota, 
\end{aligned}
\end{equation}
with
\begin{equation}
\mathcal{O}_1 =
\begin{pmatrix}
\sin \lambda \cos \psi - \cos \lambda \sin \theta \sin \psi & -\sin \lambda \sin \psi - \cos \lambda \sin \theta \cos \psi & -\cos \lambda \cos \theta  \\
     -\cos \lambda \cos \psi - \sin \lambda \sin \theta \sin \psi & \cos \lambda \sin \psi - \sin \lambda \sin \theta \cos \psi & -\sin \lambda \cos \theta  \\
         \cos \theta \sin \psi & \cos \theta \cos \psi & -\sin \theta 
\end{pmatrix},
\end{equation}
where $\psi$ is the polarization angle. The response to the GW polarization p in the link from S/C$i$ to $j$ will be
\begin{equation}
\begin{aligned}
y^{h}_{\mathrm{p}, ij} (f) =&  \frac{ \hat{n}_{ij} \cdot {\mathrm{ e_p}} \cdot \hat{n}_{ij} }{2 (1 - \hat{n}_{ij} \cdot \hat{k} ) } 
 \times \left[  \exp( 2 \pi i f (L_{ij} + \hat{k} \cdot p_i ) ) -  \exp( 2 \pi i f  \hat{k} \cdot p_j )  \right] ,
\end{aligned}
\end{equation}
\end{widetext}
where $\hat{n}_{ij}$ is the unit vector from S/C$i$ to $j$, $L_{ij}$ is the arm length from S/C$i$ to $j$, $p_i$ is the position of the S/C$i$ in the solar-system barycentric (SSB) ecliptic coordinates.

The response of a TDI combination for a specific polarization p in the frequency domain will be simplified by summing up the responses in the time shift single links. For instance, the response in the X channel could be described by
\begin{equation} \label{eq:resp_X_FD}
\begin{aligned}
 F_{ \rm X,p} (f) =& (-\Delta_{21} + \Delta_{21}  \Delta_{13}  \Delta_{31})  y^{h}_\mathrm{p,12} \\
         & + (-1 + \Delta_{13}  \Delta_{31} )  y^{h}_\mathrm{p,21}  \\
         & + (\Delta_{31} - \Delta_{31}  \Delta_{12}  \Delta_{21})  y^{h}_\mathrm{p,13} \\
         & + ( 1 - \Delta_{12}  \Delta_{21} )  y^{h}_\mathrm{p,31}, \\
\end{aligned}
\end{equation}
where $\Delta_{ij} = \exp(2 \pi i f L_{ij})$. The GW responses in the optimal A, E, and T channels are obtained by applying Eq. \eqref{eq:optimalTDI} straightforwardly.
One polarization of GW waveform in a TDI channel could be expressed as $\tilde{h}_\mathrm{TDI, p} =  F_\mathrm{TDI, p} \tilde{h}$ as it will be shown in Fig. \ref{fig:LISA_sources_sensitivity}, where the $\tilde{h}$ is the {\it intrinsic} GW waveform in the frequency domain. By using Eq. \ref{ppe6waveform}, The alternative GW waveform with six polarizations in one TDI channels could be modified as \cite{yunes2012PhRvD..86b2004C}
\begin{equation} \label{eq:h_FD_ppE}
\begin{aligned}
\tilde{h}_\mathrm{ppE, TDI} (f) = &  \left[ ( F_{+} + F_{\times} ) ( 1 + c \beta u^{b+5}_2 ) + \alpha_\mathrm{b} F_\mathrm{b} + \alpha_\mathrm{L} F_\mathrm{L} \right. \\
& \left. + \alpha_\mathrm{x} F_\mathrm{x} + \alpha_\mathrm{y} F_\mathrm{y}   \right] \tilde{h}_\mathrm{GR} \ e^{ 2 i \beta u^b_2} ,
\end{aligned}
\end{equation}
where $\tilde{h}_\mathrm{GR} $ is the intrinsic GW waveform from GR which described by the approximant IMRPhenomPv2 \cite{Khan:2015jqa} in our calculations. The $\beta, b, \alpha_\mathrm{b}, \alpha_\mathrm{L}, \alpha_\mathrm{x}$ and $\alpha_\mathrm{y} $ are the six ppE parameters  to be determined. 

\subsection{The average sensitivities of LISA to the polarizations}

Considering the various response in TDI channels, we evaluate the average sensitivities of the LISA and LISA-TAIJI network to the six polarization modes at first. Following the method we used in \citet{Wang:2020cpq}, $10^5$ sources are simulated randomly which are located over the sky and polarization at each frequency. The response of one TDI channels to a polarization mode is calculated by using the Eq. \eqref{eq:optimalTDI}-\eqref{eq:resp_X_FD} with an optimal inclination, (for instance, inclination $\iota = 0$ yields the maximum amplitude for tensor polarization, and $\iota=\pi/2$ yields the strongest GW for scalar polarizations).
The median responses of joint A+E+T channels over sky and polarization angle are chosen to represent the average capacity of one mission to a specific GW polarization mode. And the response of LISA and TAIJI to a source are calculated simultaneously. 

The acceleration noise and optical path noise are considered to evaluate the sensitivity of LISA/TAIJI. The noise budgets are from the updated upper limit of their noise requirements \cite{2017arXiv170200786A,Luo:2020}. The acceleration noise $S_{\rm acc}$ requirements are assumed to be the same for both LISA and TAIJI,
\begin{equation}
 S^{1/2}_{\rm acc} = 3 \times 10^{-15} \frac{\rm m/s^2}{\sqrt{\rm Hz}} \sqrt{1 + \left(\frac{0.4 {\rm mHz}}{f} \right)^2 }  \sqrt{1 + \left(\frac{f}{8 {\rm mHz}} \right)^4 }.
\end{equation}
And the optical path noises $S_{\rm op}$ requirement for two missions are slightly different which are
\begin{equation}
\begin{aligned}
 S^{1/2}_{\rm op, LISA} & = 10 \times 10^{-12} \frac{\rm m}{\sqrt{\rm Hz}} \sqrt{1 + \left(\frac{2 {\rm mHz}}{f} \right)^4 },  \\
S^{1/2}_{\rm op, TAIJI} & = 8 \times 10^{-12} \frac{\rm m}{\sqrt{\rm Hz}} \sqrt{1 + \left(\frac{2 {\rm mHz}}{f} \right)^4 }.
 \end{aligned}
\end{equation}
The combined noise PSDs of the TDI channels are calculated by implementing the algorithm in \cite{Wang:2020fwa,Wang:2020cpq}.

\begin{figure}[htb]
\includegraphics[width=0.48\textwidth]{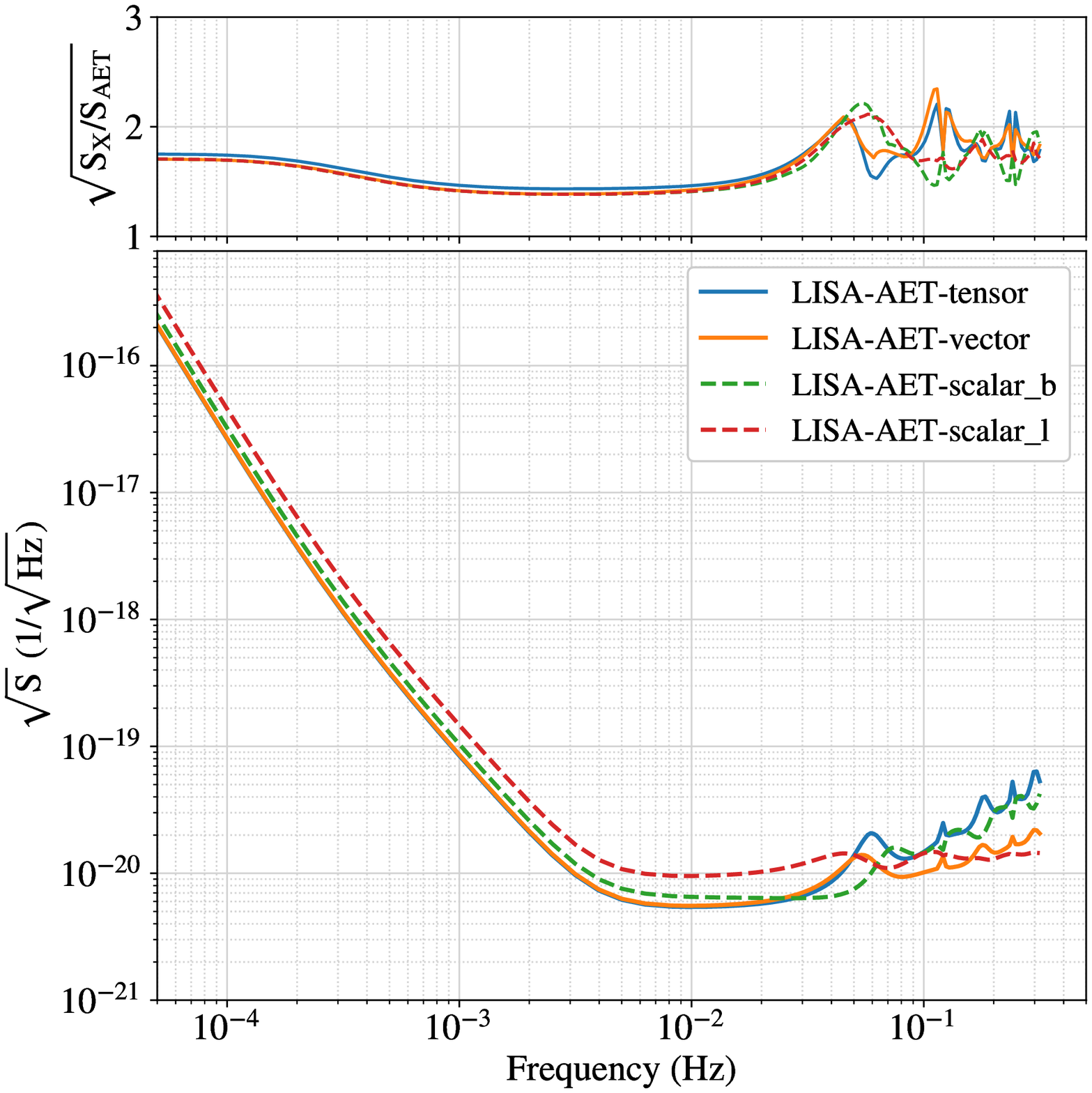} 
\includegraphics[width=0.48\textwidth]{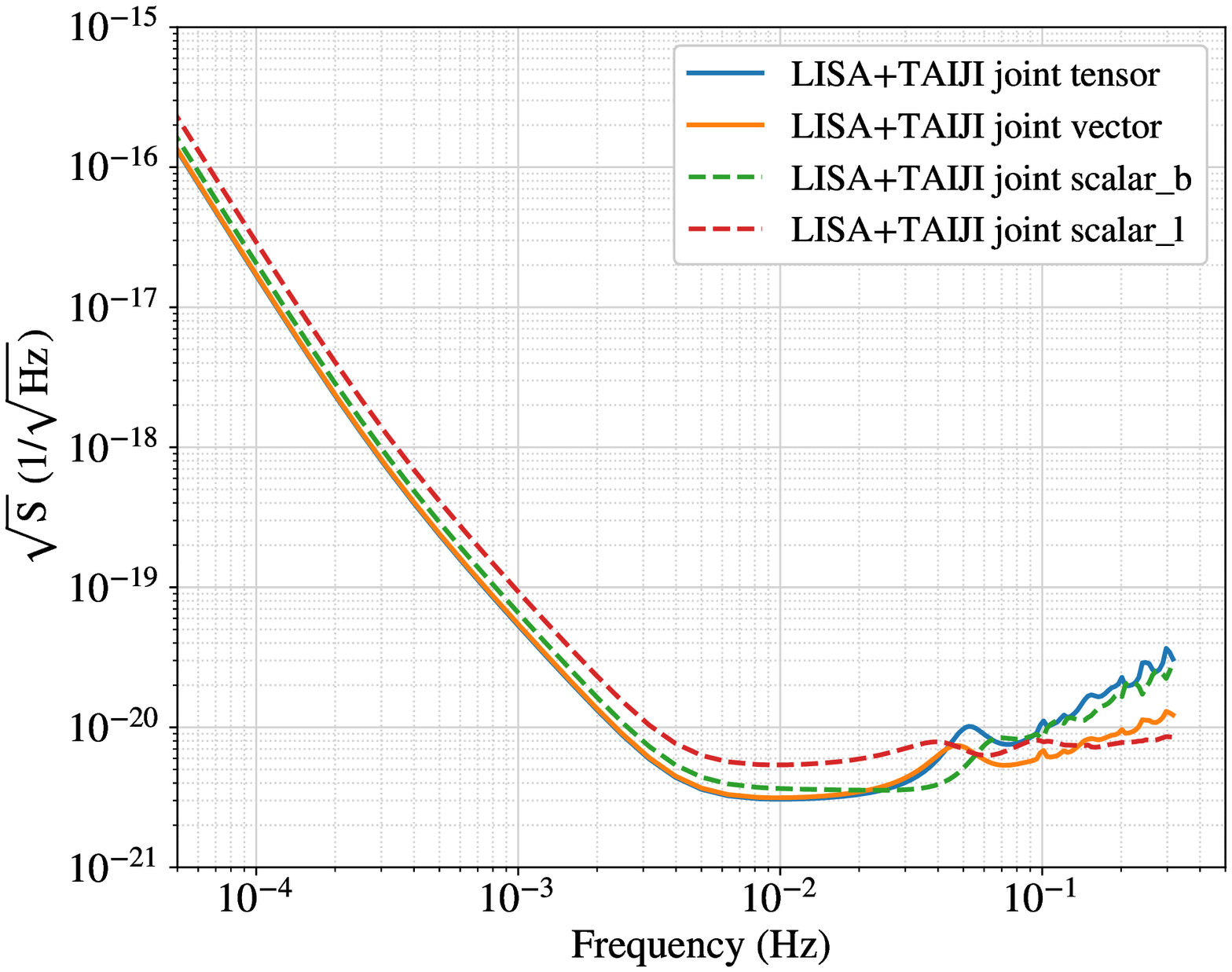} 
\caption{\label{fig:LISA_polar_sensitivity} The average sensitivities of LISA mission (upper panel) and joint LISA-TAIJI network (lower panel) to the different polarization modes at the optimal inclination angles. The upper plot of the upper panel shows the joint LISA A+E+T channel would improve the sensitivity by a factor of $\sqrt{2}$ to 2 compared to its fiducial Michelson X channel. The lower panel shows the joint LISA-TAIJI network can improve the sensitivities by a factor of $\geq \sqrt{2}$ than the single LISA mission.}
\end{figure}

The average sensitivities of the LISA's A+E+T channel and joint LISA-TAIJI network to a polarization mode p are obtained by, 
\begin{align}
\mathrm{S}_{\rm LISA, p} &= \left( \sum_{\rm A,E,T} \frac{|F_{\rm TDI, p}|^2}{ \mathrm{S}_{\rm n, TDI}} \right)^{-1}, \label{eq:S_LISA} \\
\mathrm{S}_{\rm joint, p} &= \left( \sum^{\rm TAIJI}_{\rm LISA} \sum_{\rm A,E,T} \frac{|F_{\rm TDI, p}|^2}{ \mathrm{S}_{\rm n, TDI}} \right)^{-1}.
\end{align}
The average sensitivities of the LISA mission for different polarization modes are shown in the upper panel of Fig. \ref{fig:LISA_polar_sensitivity}. The upper plot of the upper panel shows the increase of joint A+E+T sensitivity compared to the fiducial Michelson X channel. As we expatiated in \citet{Wang:2020fwa}, the joint A+E+T channels will improve the sensitivity by a factor of $\sqrt{2}$ to 2 times than X single channel. The joint LISA-TAIJI observation can further improve the sensitivity of LISA by a factor of $\geq \sqrt{2}$ as shown in the lower panel of Fig. \ref{fig:LISA_polar_sensitivity}. We can also notice that the sensitivity for vector mode and the longitudinal mode not quickly decline as the tensor mode, and this should be due to the higher response at high frequency band in the TDI for these polarization modes as discussed in \cite{Tinto:2010hz,Zhang:2019oet}.

The sensitivities for alternative polarizations in Fig. \ref{fig:LISA_polar_sensitivity} are calculated by assuming the ppE parameter $\alpha_i = 1$ in Eq. \eqref{eq:h_FD_ppE} and an optimal inclination $\iota$ in Eq. \eqref{eq:polarizations-response}. The sensitivities could be scaled by the tuned factors.

\section{Constraining ppE parameters from SMBH binary Coalescence} \label{sec:SMBH_results}

\subsection{Source selections}

Following our previous work \cite{Wang:2020a}, we choose the SMBH binaries with mass ratio $q= 1/3 $ at redshift $z=2$ to examine the detectability of the LISA-TAIJI network and compare the results to a single LISA mission. Two masses setups are employed which are source1 ($m_1 = 10^5 M_{\odot}, m_2 = 3.3 \times 10^4 M_{\odot} $) and source2 ($m_1 = 10^6 M_{\odot}, m_2 = 3.3 \times 10^5 M_{\odot} $). 
Another motivation for this selection is that these two sources could be well sky localized by the two detector network as we studied in \cite{Wang:2020a}. Therefore, an optimistic scenario would be assumed that the source location (direction and distance) could be determined by multi-messenger observation, and the known source location may improve the achieved results. 

\begin{figure}[htb]
\includegraphics[width=0.48\textwidth]{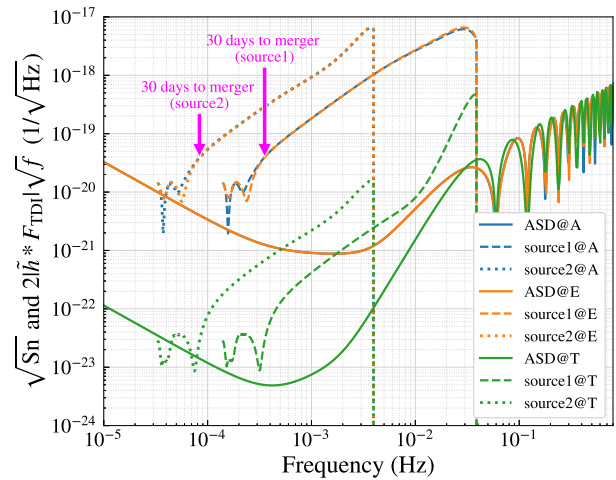} 
\caption{\label{fig:LISA_sources_sensitivity} The redshifted GW amplitudes $ 2 \sqrt{f} |\tilde{h}_{\rm GR} \ast F_{\rm TDI,p} |$ of the selected sources and ASDs in the optimal TDI channels for one year evolution before coalescence. The GW amplitudes include the TDI response function $F_\mathrm{TDI,p}$ for the sources with geometric angles $(\theta=\pi/10, \iota=0.55~ \mathrm{rad}, \psi = \pi/3)$. The ASDs of the TDI noises include the acceleration noises and optical path noises. The source1@TDI indicates GW amplitude of the source1 ($m_1 = 10^5 M_{\odot}, q=1/3, z=2$) in the TDI channels, and source2@TDI indicates the source2 ($m_1 = 10^6 M_{\odot}, q=1/3, z=2$). }
\end{figure}

The redshifted GW amplitudes of two sources in the selected TDI channels and the ASDs of the channels are shown in Fig. \ref{fig:LISA_sources_sensitivity}. The amplitudes incorporate the response function of the TDI channels $ 2 \sqrt{f} |\tilde{h}_{\rm GR} \ast F_{\rm TDI,p} | $ for the specific source parameters $(\theta=\pi/10, \iota=0.55 ~\mathrm{rad}, \psi = \pi/3)$ through the frequency band in the last one year of coalescing. The ASDs of the TDI channels are the noise level from the acceleration and optical-path noises. The ASDs of A and E channels are identical, while the amplitudes of GW signals in their channels are different. The frequencies at 30 days before the coalescences are annotated in the plot, and the SNRs from the last 30 days are expected to be dominant for the detections.
As we previously studied in \cite{Wang:2020fwa,Wang:2020cpq}, the location around the ecliptic latitude $ \theta = 18^\circ$ would be an optimal choice for the average response, and the longitude of sources is coordinated with the positions of the LISA and TAIJI to have an optimal response. The inclination of the sources is one of the key factors which relates to the cadence of the different polarization amplitudes. By presuming the polarization modes beyond GR are much less significant than $+/\times$ polarizations from GW, the inclination angle is set to be $\iota = 0.55 ~\mathrm{rad}$ which is close to the favored angle of the detections as shown in the upper plot of Fig. \ref{fig:iota_distribution_SNR_curves} \cite{Schutz:2011tw}. And we will also perform the investigations varying with the inclination in Section \ref{subsec:results_of_iota}. The polarization angle is set to be $\psi = \pi/3$.

For the source1 ($m_1 = 10^5\ M_{\odot}, q=1/3$), the GW frequency evolution during the one year to merger will change from 0.14 mHz to 40 mHz, and the corresponding $u^3_2 = \pi \mathcal{M} f$ value changes from [0.0003, 0.09]. And the GW frequency from the binary ($m_1 = 10^6 M_{\odot}, q=1/3$) will start from 0.033 mHz to 4 mHz, and the range of $u^3_2$ is  [0.00075, 0.09]. As estimated in \citet{Cornish:2011ys}, the bounds limits of $\beta$ at given $b$ is expected to be inversed proportional to the SNR and the range of $u^b_2$.

\subsection{Antenna patterns for polarizations}

The response of a GW interferometer to the GW signals changes with the source locations and orientations. The antenna patterns of a ground-based interferometer for the alternative GW polarization modes have been plotted in \cite{Nishizawa2010PhRvD,Isi2015PhRvD}. To illustrate the antenna pattern of a LISA-like mission, the joint responses of the A, E, and T TDI channels to each polarization in the detector frame are shown in Appendix \ref{sec:appendix} Fig. \ref{fig:antenna_polarization}.  As the Fig. \ref{fig:antenna_polarization} shown, the most sensitive directions for the tensor modes are the normal/polar directions with respect to the interferometer plane, while the most sensitive directions for the scalar and vector x polarization are the equatorial directions. The optimal direction for the vector y polarization observation is the direction $\pi/4$ with respect to the formation plane.

\begin{figure*}[htb]
\includegraphics[width=0.46\textwidth]{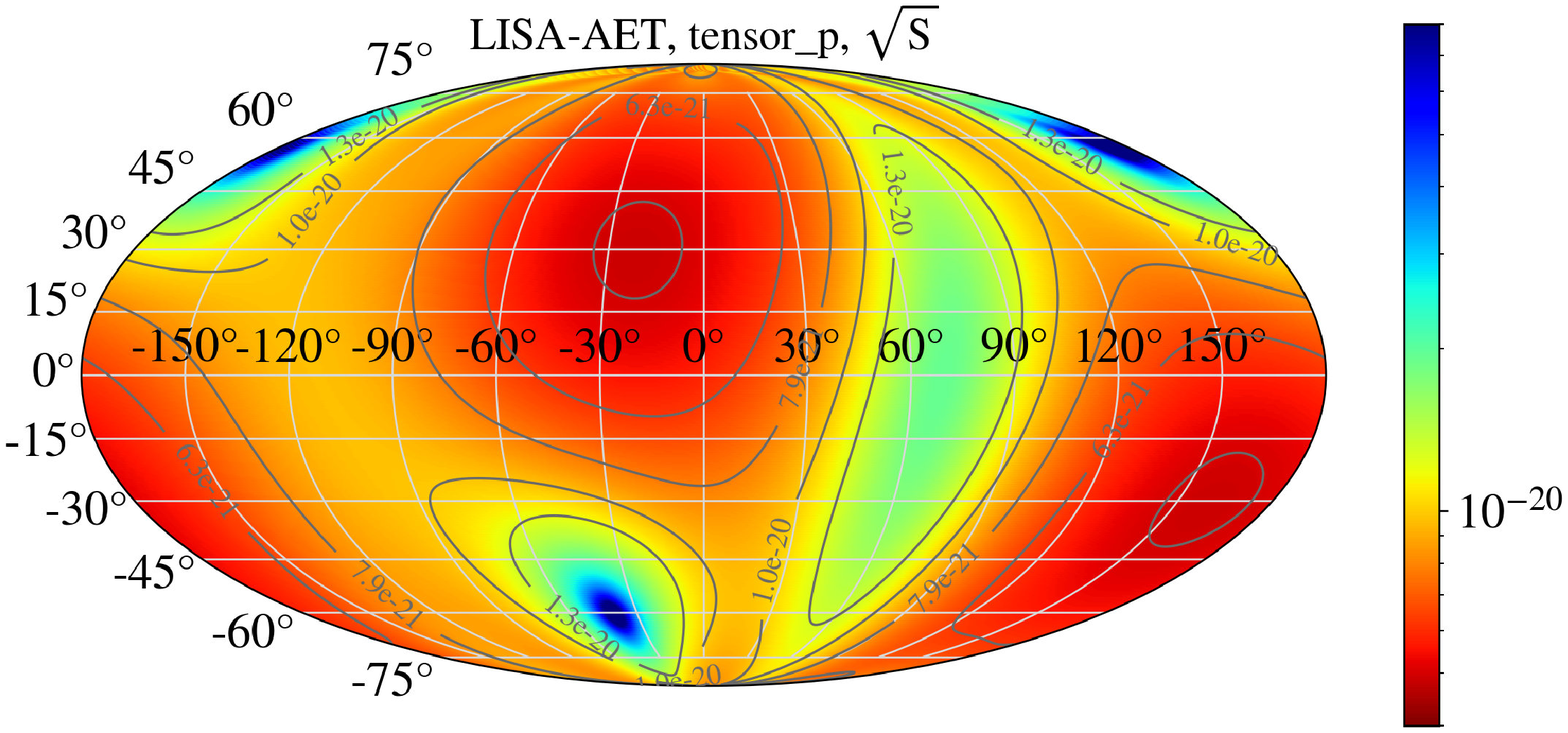}
\includegraphics[width=0.46\textwidth]{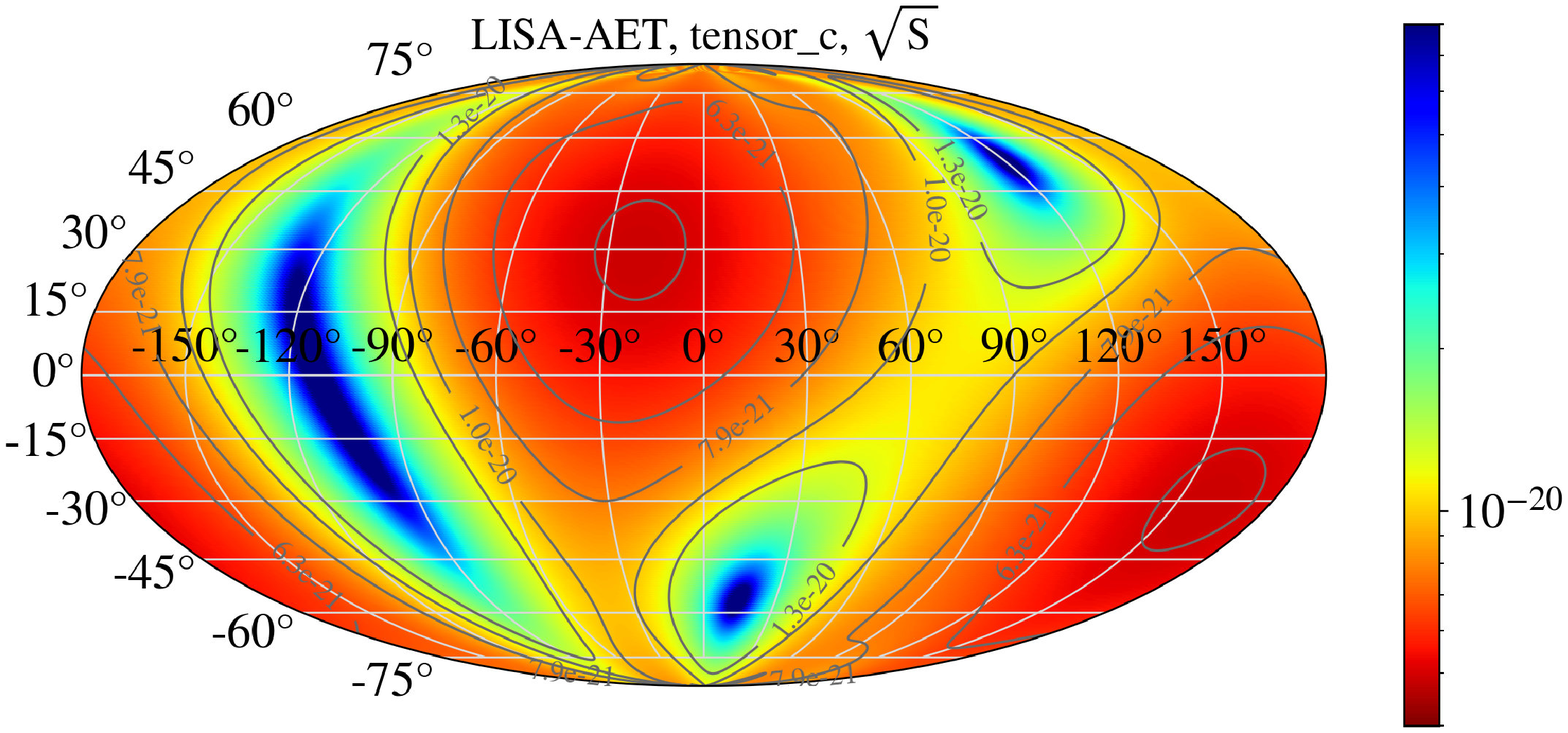}
\includegraphics[width=0.46\textwidth]{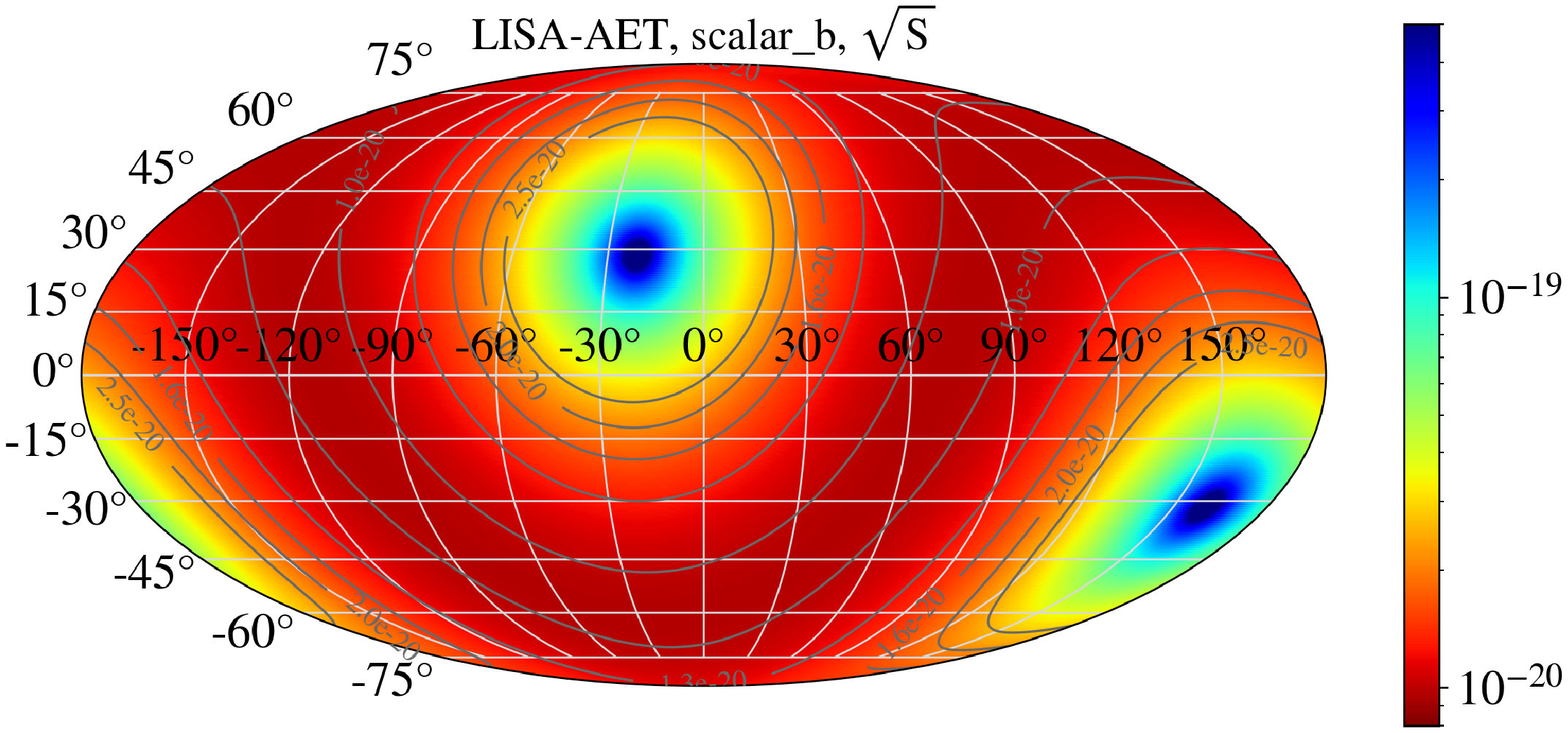}
\includegraphics[width=0.46\textwidth]{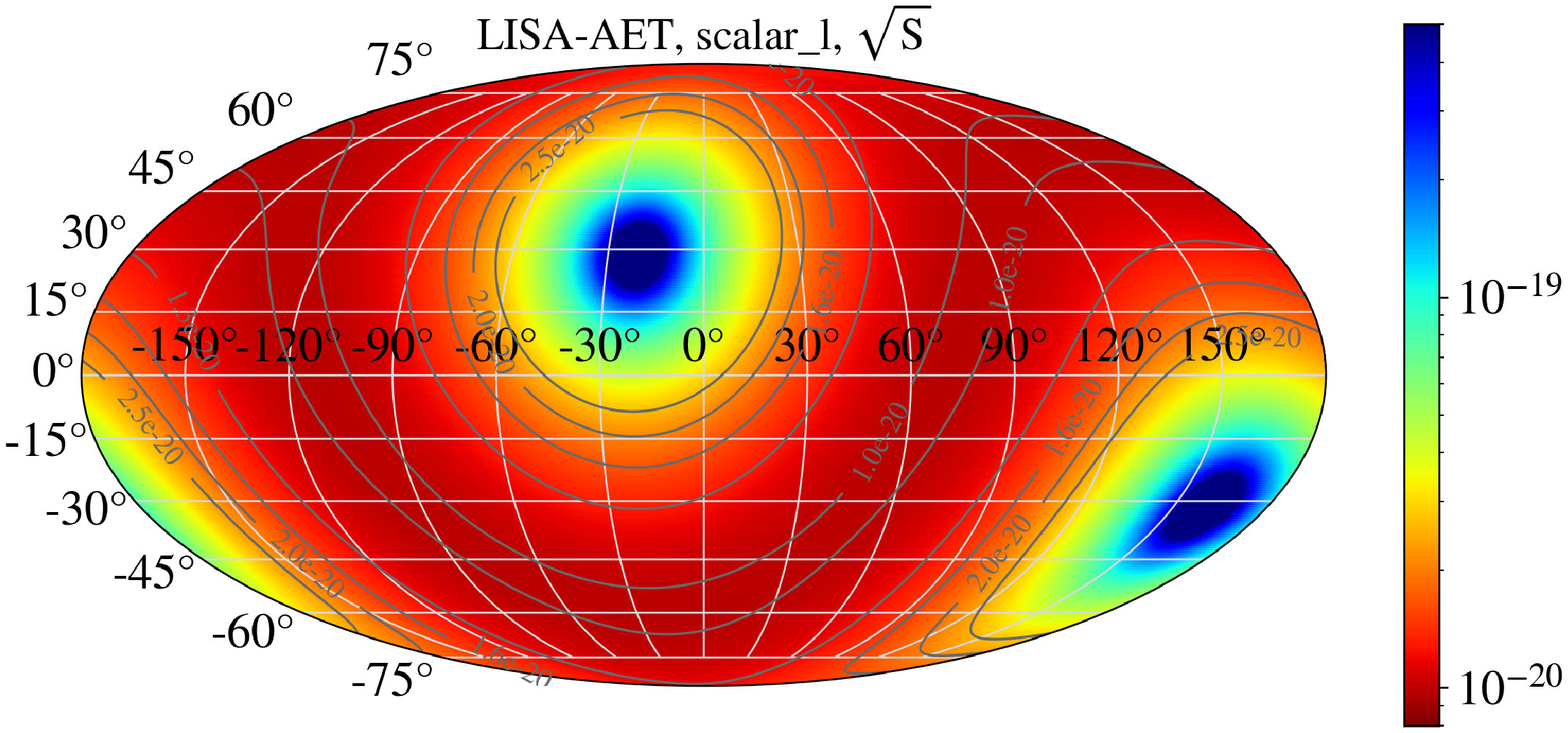}
\includegraphics[width=0.46\textwidth]{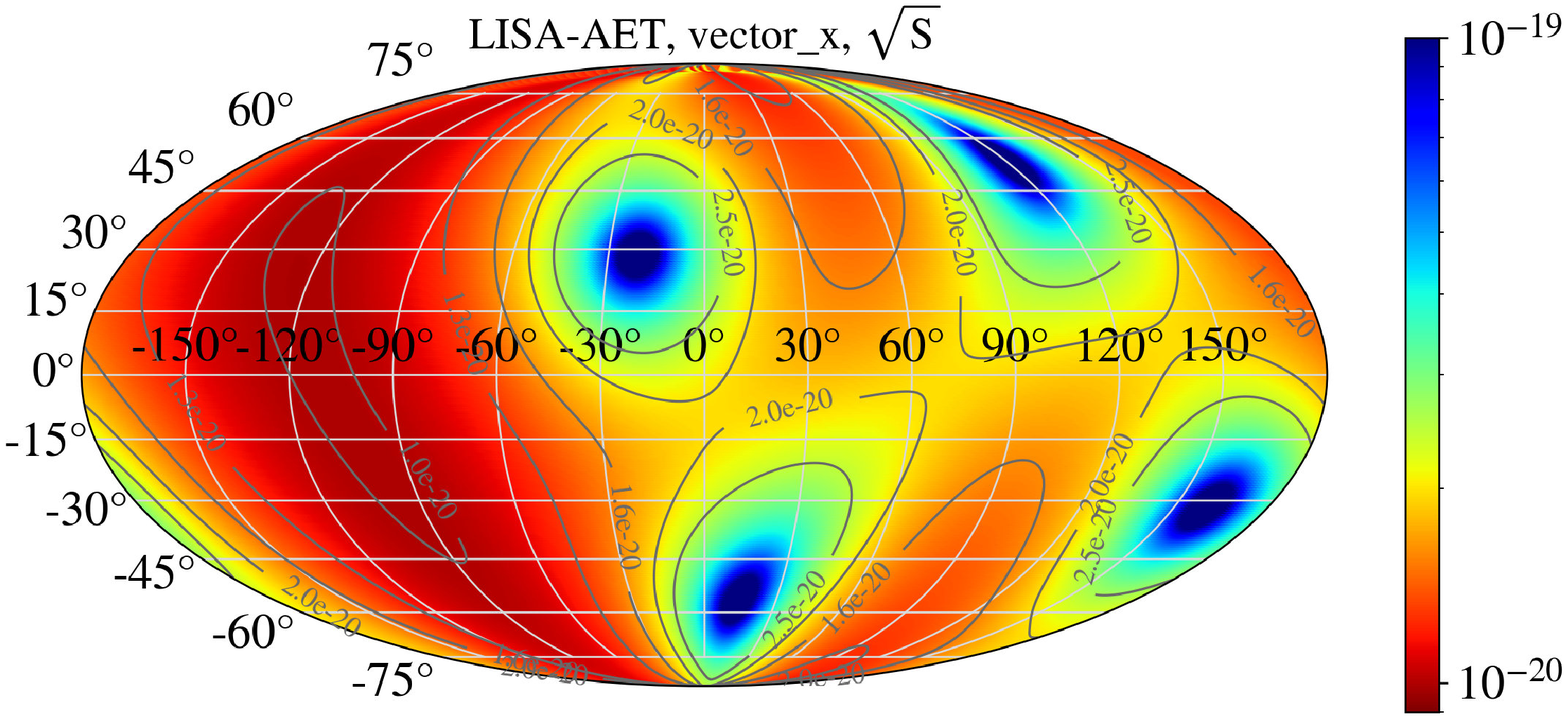} 
\includegraphics[width=0.46\textwidth]{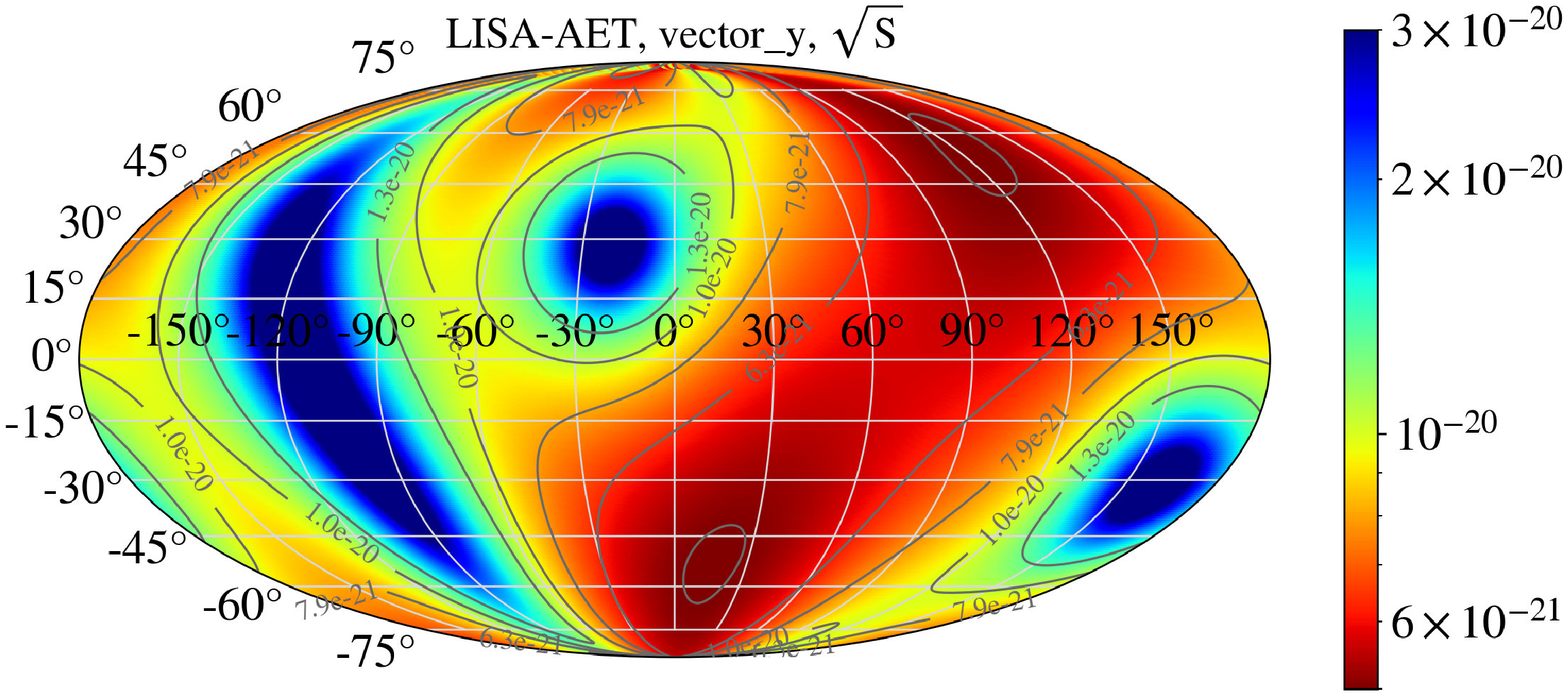} 
\caption{\label{fig:molllweide_polarization} The instantaneous sensitivities of LISA joint A+E+T channels for various polarization modes at 10 mHz. The $\psi$ is set to be $\pi/3$, and inclination $\iota$ is optimal for each polarizations.}
\end{figure*}

As aforementioned, the S/C formation plane of a LISA-like mission has 60$^{\circ}$ inclination angle with respect to the ecliptic plane. Considering the detector's orbital motion, the SSB coordinates are employed to incorporate the modulations with the relative positions and orientation changes between the interferometer and the GW sources.
The instantaneous sensitivities of LISA for the different polarization modes are shown in Fig. \ref{fig:molllweide_polarization}. The sensitivity is calculated by using Eq. \eqref{eq:S_LISA} at 10 mHz for the $\psi = \pi/3$ and optimal inclinations $\iota$ for each polarization mode. As we can see in two plots of the upper panel, the most sensitive direction for the tensor modes observation is around $\pm 30^\circ$ ecliptic latitude facing by the S/C triangular formation. For scalar polarization modes, the most sensitive directions of the tensor modes are the most insensitive directions. For vector modes, besides the insensitive directions to the normal directions of the S/C formation plane, there are some other unresponsive directions. This antenna pattern also will change with the geometric angles ($\psi$ and $\iota$), the GW frequency, and the time. With a $40^\circ$ separation from LISA as Fig. \ref{fig:LISA_TAIJI} shown, the TAIJI mission is expected to have similar antenna pattern with $\sim40^\circ$ spatial shifted along the ecliptic latitude. The joint LISA and TAIJI network will improve their sky coverage and enhance their detectability as we will see in the following subsections.

In this work, the coalescing SMBH binaries are selected to investigate the polarization observations, and the GW signals observations are mainly concentrated in the last one month indicated in Fig. \ref{fig:LISA_sources_sensitivity}. On the other side, the observations of dominant tensor polarizations have the inversed favored sky direction against other alternative polarizations. It would be a trade-off for the selections of the source location and merger time. When the observation simulation is beneficial to the tensor modes, the SNR could increase and ppE parameters $\beta$ and $b$ could be well constrained, and other polarization modes could be poorly observed and parameters $\alpha_i$ may be underestimated, vice versa. 
In this investigation, the sources are selected to make their mergers happen around detectors' sensitive directions for the tensor mode.

\subsection{Fisher information method}

The Fisher information matrix (FIM) is applied in this investigation to determine the uncertainty of parameter estimation from GW observation as used in \cite[and references therein]{1994PhRvD..49.2658C,Cutler:1997ta,Vallisneri:2007ev,Kuns:2019upi,Wang:2020a}. 
The FIM from the single LISA mission is obtained by summing up the three optimal channels (A, E, and T), and the joint FIM of the LISA-TAIJI network is achieved by summing up the FIM from each mission,
\begin{equation}
\Gamma_{ij}  = \sum^{\mathrm{TAIJI}}_{\rm LISA} \sum_{\rm A,E,T}  \left( \frac{\partial \tilde{h}_\mathrm{ppE,TDI} }{ \partial {\xi_i} } \bigg\rvert \frac{\partial \tilde{h}_\mathrm{ppE,TDI} }{ \partial {\xi_j} }  \right),
\end{equation}
with
\begin{equation}
\left( g | h\right)_{\rm TDI} = 4 \mathrm{Re} \int^{\infty}_0 \frac{g^{\ast} (f) h(f)}{S_{\rm TDI} (f) } \mathrm{d} f ,
\end{equation}
where $\tilde{h}_\mathrm{ppE,TDI}$ is the frequency domain GW waveform with all polarization modes as described by Eq. \ref{eq:h_FD_ppE}, $\xi_i$ is the $i$-th parameter to the determined, and $S_{\rm TDI} (f)$ is the noise PSD of one TDI channel from LISA or TAIJI. 

In this investigation, 15 parameters are considered to describe the GW signal from a binary system which are ecliptic longitude and latitude $(\lambda, \theta)$, polarization angle $\psi$, inclination $\iota$, luminosity distance $D$, the coalescence time and phase $(t_c, \phi_c)$, the total mass of binary $M$ and mass ratio $q$, and the six ppE parameters ($\beta, b, \alpha_\mathrm{b}, \alpha_\mathrm{L}, \alpha_\mathrm{x}, \alpha_\mathrm{x} $).
Two scenarios are considered to implement the FIM calculations, the first one is that the location of the source is unknown and the FIM is calculated for full 15 parameters, the second case is that the location of GW source is known from other associated observation and FIM is calculated for 12 parameters excluding the three parameters, $(\lambda, \theta, D)$. 

The variance-covariance matrix of the parameters could be obtained by
\begin{equation}
\begin{aligned}
 \left\langle \Delta \xi_i \Delta \xi_j  \right\rangle = \left( \Gamma^{-1} \right)_{ij} + \mathcal{O}({\rho}^{-1}) 
\end{aligned}
\end{equation}
The standard deviations $\sigma_i$ and correlation coefficients $\sigma_{ij}$ of the parameters for the high SNR $\rho \gg 1$ will be
\begin{equation}
\begin{aligned}
\sigma_{i} & \simeq \sqrt{ \left( \Gamma^{-1} \right)_{ii}  }, \\
\sigma_{ij} & = \frac{\mathrm{cov} (\xi_i, \xi_j) }{\sigma_i \sigma_j} \simeq \frac{ \left( \Gamma^{-1} \right)_{ij} }{\sigma_i \sigma_j}.
\end{aligned}
\end{equation}
We focus on the ppE parameters determinations from LISA and the improvements from LISA-TAIJI joint observations in this work.

\subsection{Results with varying inclination} \label{subsec:results_of_iota}

\begin{figure}[htb]
\includegraphics[width=0.48\textwidth]{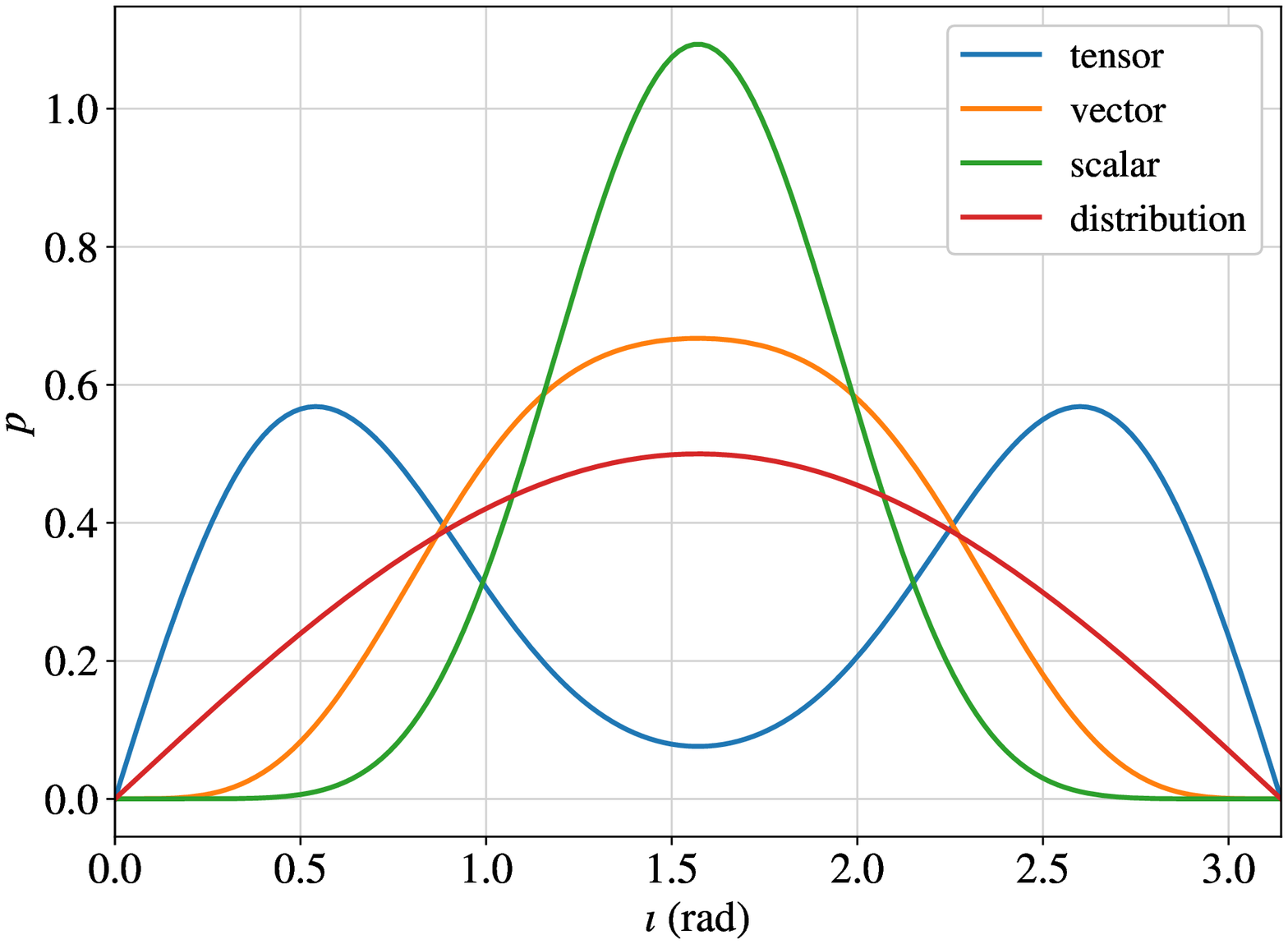} 
\includegraphics[width=0.48\textwidth]{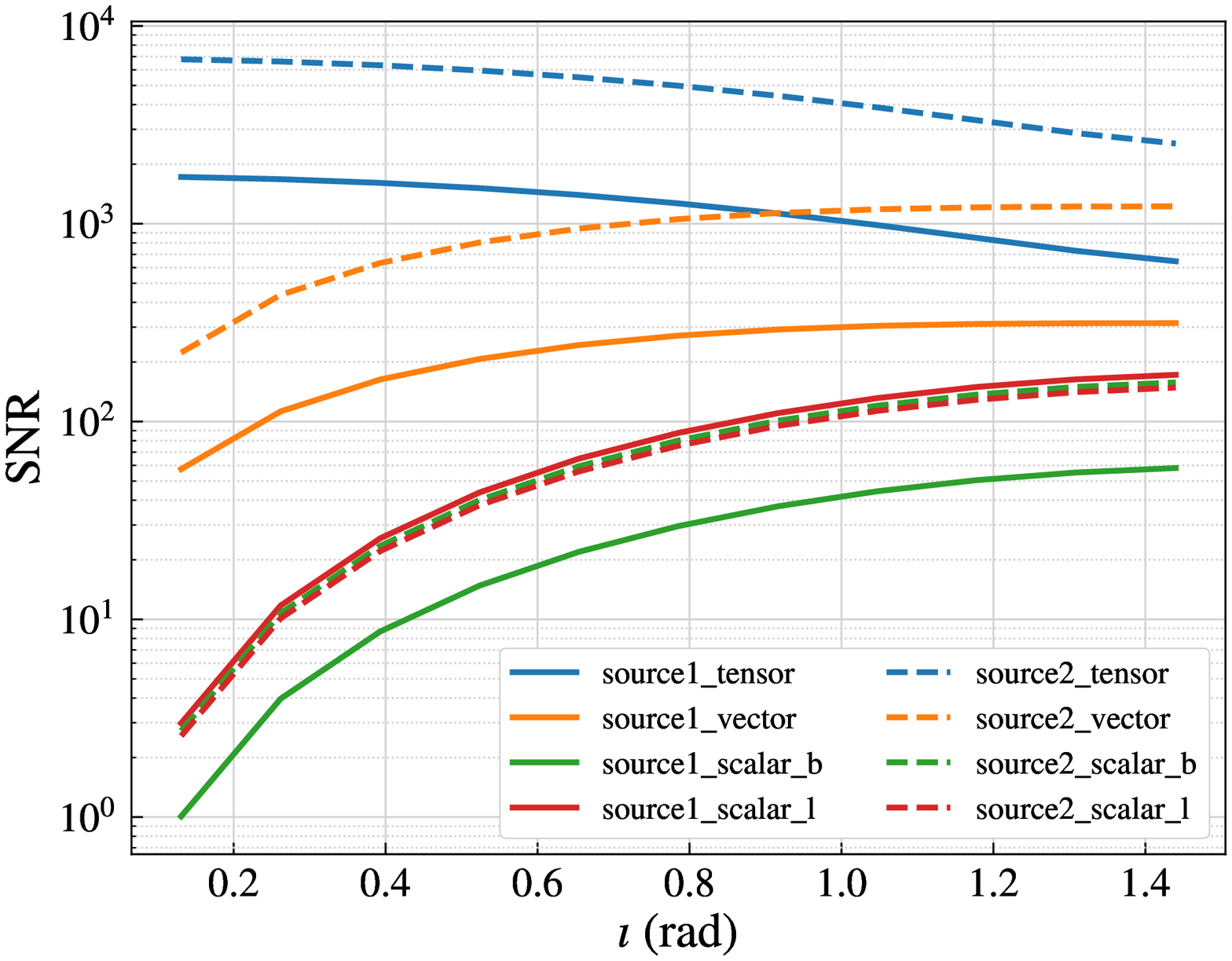} 
\caption{\label{fig:iota_distribution_SNR_curves} The distributions of the inclination $\iota$ from expected detections considering the solo  tensor, vector, and scalar polarization modes (upper panel) and the LISA's SNR of polarizations for corresponding $\alpha_i = 1$ from two selected sources with the inclinations (lower panel). The {\it tensor} curve shows the distribution of $\iota$ from the standard GR GW detections which described by Eq. \eqref{eq:iota_dist_tensor} \cite{Schutz:2011tw}. The {\it vector} and {\it scalar} curve show the distributions of $\iota$ when the solo vector or scalar polarization GW waveform is detected as described by Eq. \ref{eq:iota_dist_vector_scalar}. The {\it distribution} curve shows the distribution of $\iota$ from geometry, $p(\iota) \propto \sin \iota $.
}
\end{figure}

We examine the detectability of the ppE parameter varying with the inclination of sources in this subsection.
The amplitude of a GW signal is modulated with the inclination $\iota$ of the binary as we can read from Eq. \eqref{eq:polarizations-response}. With only considering the two GW polarizations from GR, the distribution of inclination $\iota$ from detections is expected to be \cite{Schutz:2011tw}
\begin{equation} \label{eq:iota_dist_tensor}
p_\mathrm{tensor} (\iota) \propto ( 1+ 6 \cos^2 \iota + \cos^4 \iota )^{3/2} \sin \iota.
\end{equation}
The normalized distribution is shown by the blue curve in the upper plot of Fig. \ref{fig:iota_distribution_SNR_curves}. Similarly, if the vector polarizations or the scalar polarization is only considered, the corresponding distributions of inclination angle will be
\begin{equation} \label{eq:iota_dist_vector_scalar}
\begin{aligned}
p_\mathrm{vector} (\iota) & \propto ( \sin^2 \iota \cos^2 \iota + \sin^2 \iota )^{3/2} \sin \iota, \\
 p_\mathrm{scalar} (\iota) & \propto \sin^7 \iota.
\end{aligned}
\end{equation}
Their curves are shown in Fig. \ref{fig:iota_distribution_SNR_curves} upper plot by the orange and green curves, respectively. These distributions show the favored inclinations by the different polarizations, the most favored inclination by the tensor polarization is around $\iota = 0.55 ~\mathrm{rad}$, the distributions of $\iota$ have the peaks around $\pi/2 = 1.57$ for both vector and scalar polarizations. 

By assuming the $\alpha_i=1$ for the alternative polarizations and tensor polarizations from GR, their SNRs varying with the inclinations from two selected sources are shown in the lower plot of Fig. \ref{fig:iota_distribution_SNR_curves}. The SNR from ($\iota = \pi/2$, edge-on) will be $1/\sqrt{8}$ of SNR from ($\iota=0/\pi$, face-on/off) for the dominant tensor mode. For the inclination selection, considering the symmetry effects of the inclination in $[0, \pi/2]$ and $[\pi/2, \pi]$ range, we perform the investigations for $\iota = n \pi/24 \ (n=1 {\rm{\ to\ }} 11)$. One reason to avoid the $\iota = 0$ is the astrophysical unlikely as shown in the upper plot of Fig. \ref{fig:iota_distribution_SNR_curves}, another reason is that $\iota =0$ or $\pi/2$ will dissolve some polarizations and make the FIM singular. The ppE parameter $b$ is given different values for the different gravitational theories as shown in Table I of \cite{Cornish:2011ys}. For the first step, we pick the $b=-3$ which corresponds to the massive graviton theory \cite{Will:1997bb,Will:2004xi,Berti:2005qd,Stavridis:2009mb,Arun:2009pq,Keppel:2010qu,Yagi:2009zm}, and we investigate the other $b$ valves in the next step. The $\beta$ parameter is roughly set to be $0.01$ from the bounded result at $b=-3$ in \cite{Cornish:2011ys}. The six ppE parameters are fixed for the FIM calculations which are $(\beta$, $b$, $\alpha_b$, $\alpha_L$, $\alpha_x$, $\alpha_y) = (0.01, -3, 0,0,0,0)$.

\begin{figure*}[htb]
\includegraphics[width=0.46\textwidth]{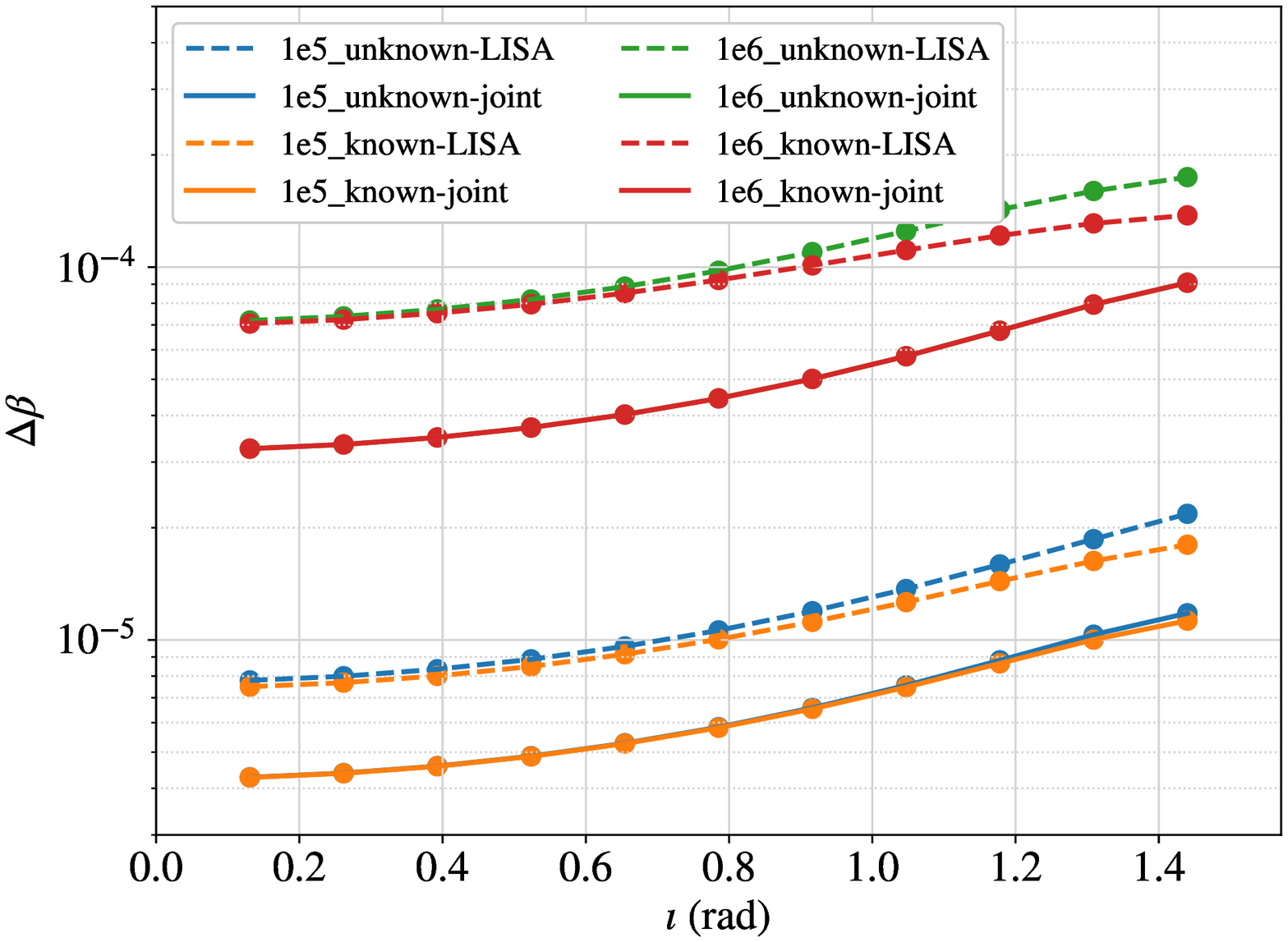} 
\includegraphics[width=0.46\textwidth]{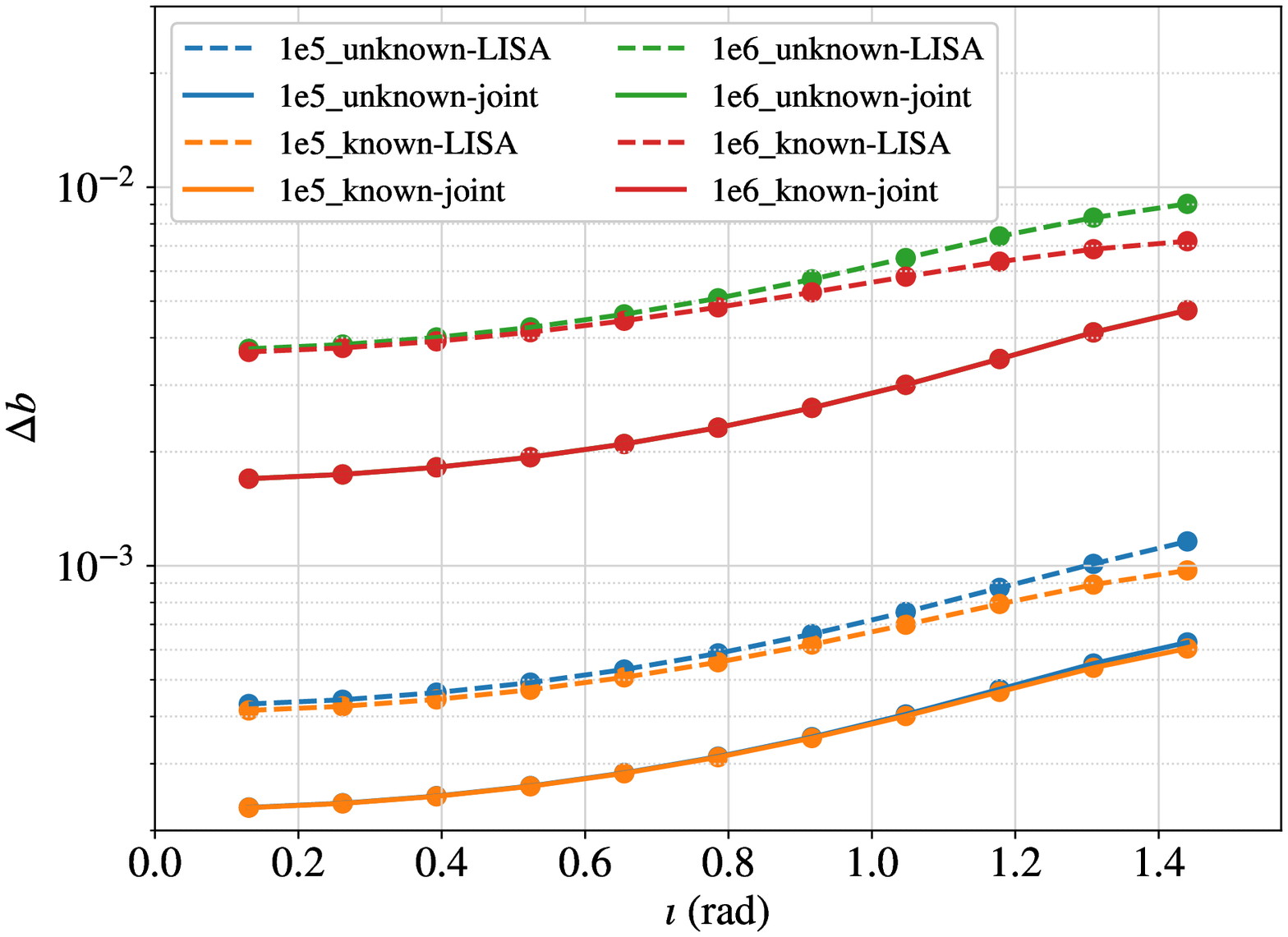} 
\includegraphics[width=0.46\textwidth]{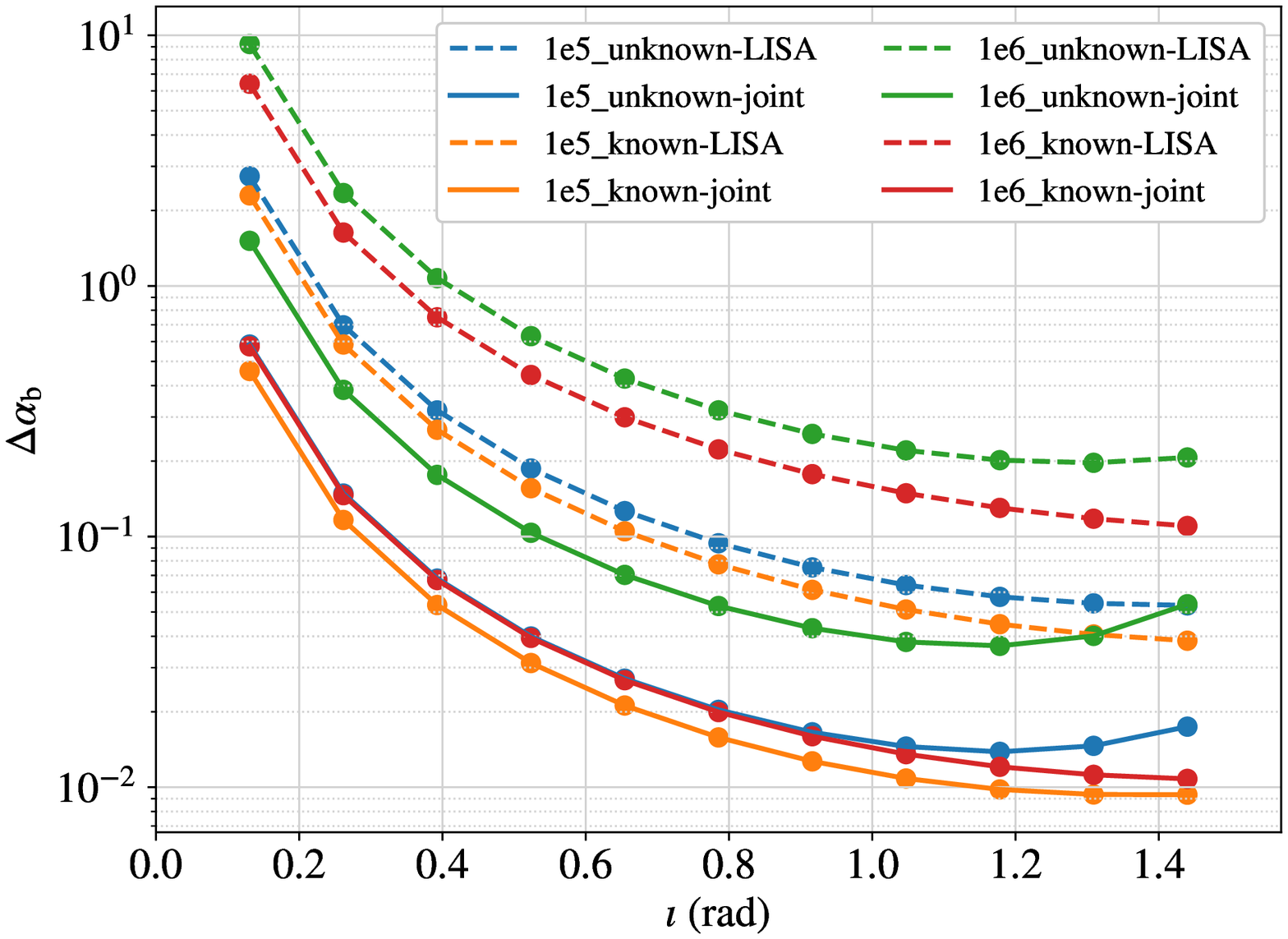} 
\includegraphics[width=0.46\textwidth]{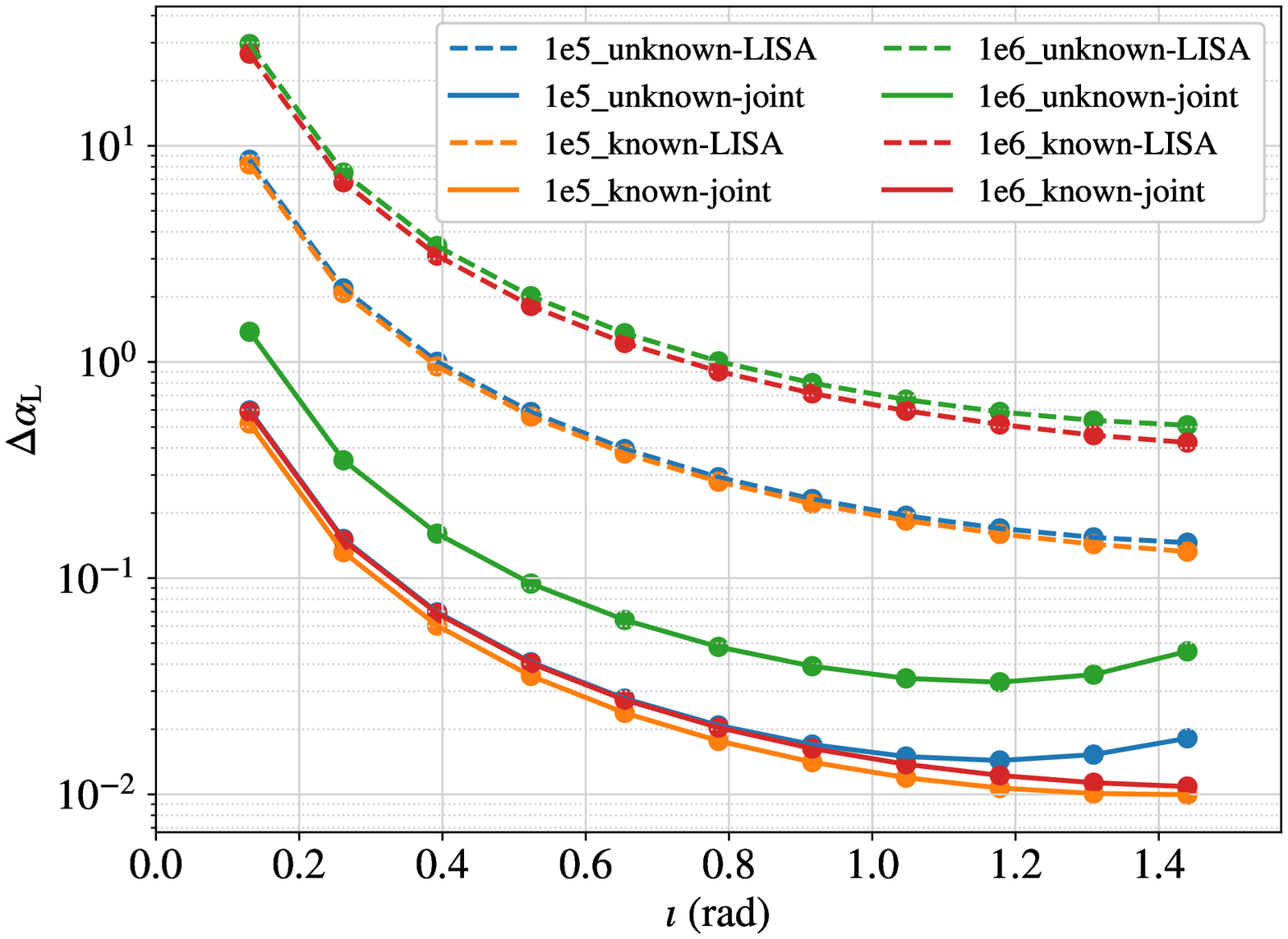} 
\includegraphics[width=0.46\textwidth]{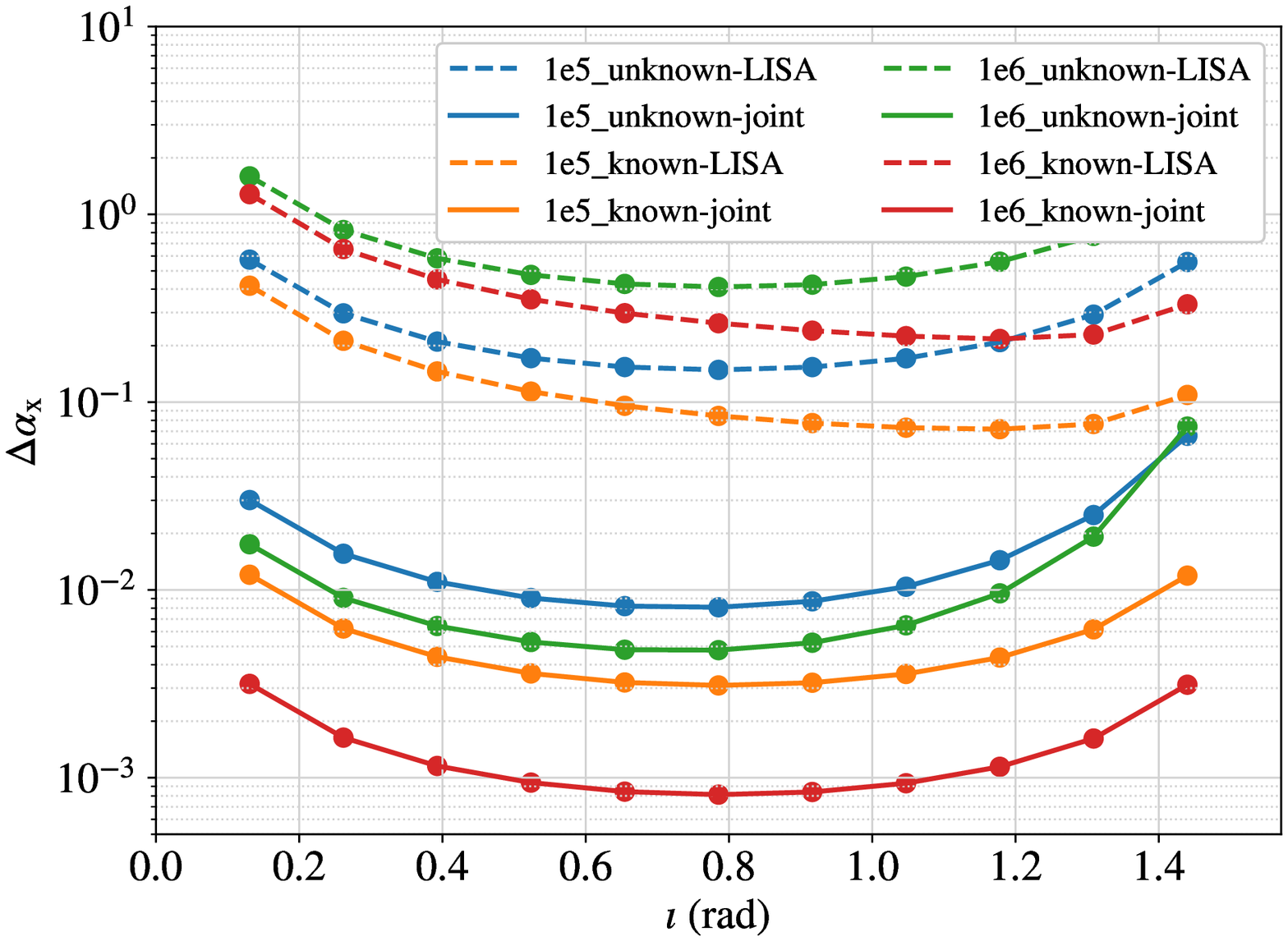} 
\includegraphics[width=0.46\textwidth]{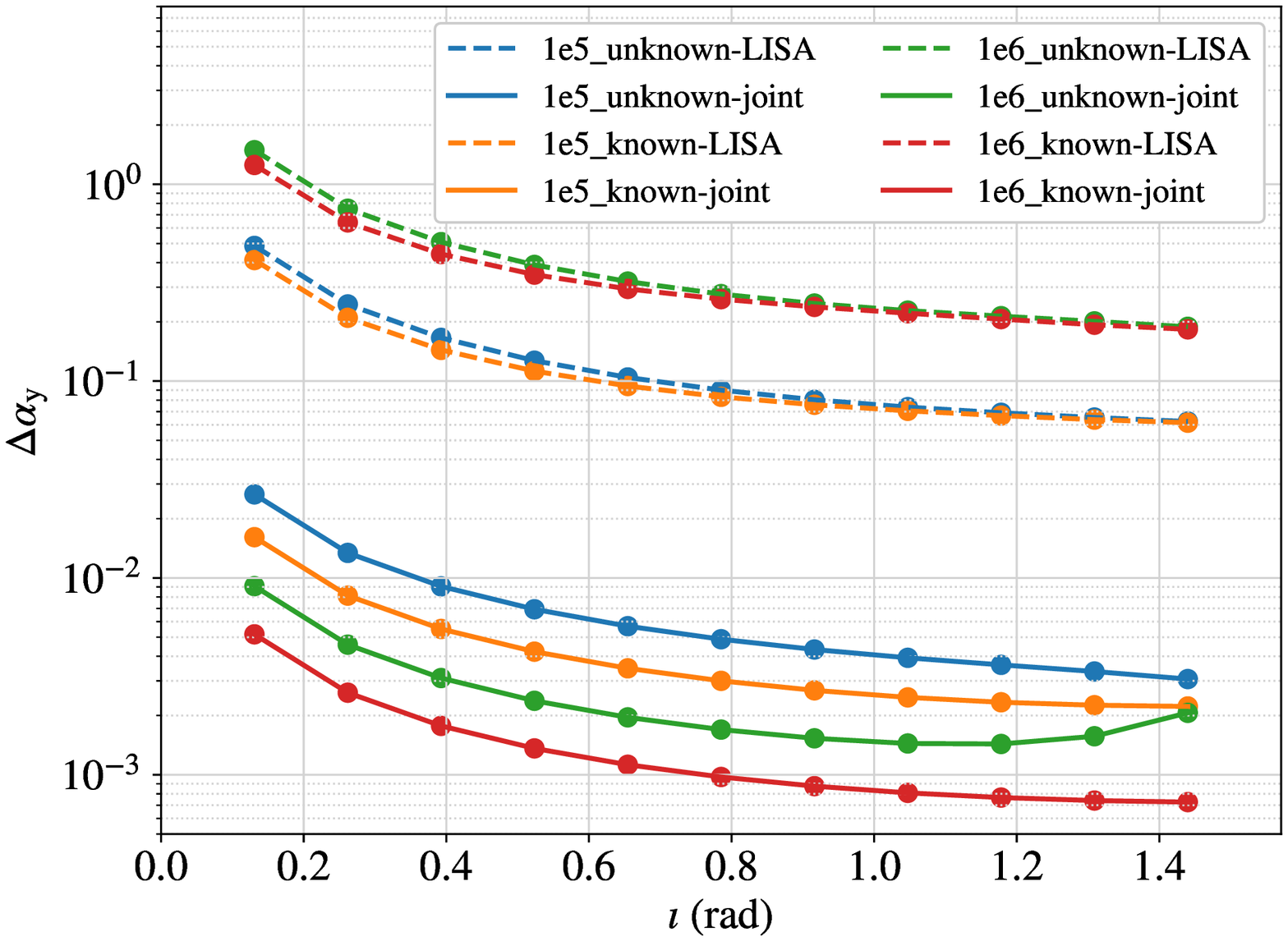} 
\caption{\label{fig:iota_vs_ppE_paras} The uncertainties of ppE parameters varying with the inclination $\iota$ from the source1 ($m_1 = 10^5 M_{\odot}, q = 1/3, z = 2$) and source2 ($m_1 = 10^6 M_{\odot}, q = 1/3, z = 2$ for ppE parameters setup $( \beta, b, \alpha_\mathrm{b}, \alpha_\mathrm{L}, \alpha_\mathrm{x}, \alpha_\mathrm{y}) = (0.01, -3, 0,0,0,0)$. Four scenarios results are shown in each plots which are 1) the result from the single LISA observation without knowing the position of the source ($m_1$\_unknown-LISA), 2) the result from joint LISA-TAIJI observation without information of  source location ($m_1$\_unknown-joint), 3) the result from the LISA observation with knowing position of the source and the FIM calculation excluding the three parameters: direction of the source ($\lambda, \theta$), and distance of the sources $D$ ($m1$\_known-LISA), and 4) the result from LISA-TAIJI joint observation with knowing the position of the source ($m_1$\_known-joint).
In the two plots in upper panel, the curves for 1e6\_unkonwn-joint and 1e6\_known-joint are overlapped.
}
\end{figure*}

The constraints on ppE parameters from the source1 ($m_1 = 10^5 M_{\odot}, q = 1/3, z = 2$) and source2 ($m_1 = 10^6 M_{\odot}, q = 1/3, z = 2$) observations for different inclination angles are shown in Fig. \ref{fig:iota_vs_ppE_paras}. The uncertainties of ppE parameters $\beta$ and $b$ get improved when the inclination approaches $0$ (or $\pi$) as shown in the two plots in the upper panel. Comparing to the LISA single detector, the joint LISA-TAIJI observation can improve the accuracy of the determination by a factor of $\sim$2 which should be the contribution of twofold SNR. When the position of sources are known and location parameters ($\lambda,~ \theta,~D$) are excluded, it only slightly improves the constraints from the single LISA observation, and does not show improvement for the LISA-TAIJI joint observation. 
Comparing the constraints from two sources, the source1 shows a better ability to measure parameters $\beta$ and $b$ than the source2. This could be due to the source1 has a relatively larger frequency range in the one year evolution, and then has a larger range of the $u_2$. The wider range of $u_2$ could improve measurements on the parameter $\beta$ and $b$ \cite{Cornish:2011ys}.
For the $\iota=0.55 ~\mathrm{rad}$, the uncertainty of $\beta$ could be constrained by the source1 in $9 \times 10^{-6}$ from LISA observation, and it could be constrained in $5 \times 10^{-6}$ by the joint observation. The uncertainty of parameter $b$ could be bound in $5 \times 10^{-4}$ by LISA from source1 observation, and be within $2.5 \times 10^{-4}$ by the LISA-TAIJI network.
In general, at any inclination case, the joint LISA-TAIJI observation could improve the $\beta$ and $b$ determinations by a factor of $\sim$2. 

The measurement uncertainties of ppE $\alpha_i$ from the two sources are shown by the middle and lower plots in Fig. \ref{fig:iota_vs_ppE_paras}. For these four parameters, the joint LISA-TAIJI network presents significant advantages. Without knowing the position of the source, the joint observation could improve the parameter measurements by more than $\sim$10 times in most of the cases except the more than $\sim$4 times improvement for the $\alpha_\mathrm{b}$. For the $\alpha_\mathrm{b}$, $\alpha_\mathrm{L}$ and $\alpha_\mathrm{y}$, their uncertainties tend to decrease with the increase of the $\iota$, and the $\alpha_\mathrm{x}$ is better measured around the $\iota = \pi/4$. We infer these tendencies from Eq. \eqref{eq:polarizations-response} that the amplitudes of scalar breathing, scalar longitudinal, and vector y polarization modes increase with the selected inclination angles, and the amplitude of vector x mode has the maximum at $\iota=\pi/4$. 

Compared to the results from two sources, the uncertainties of parameters $\alpha_\mathrm{b}$ and $\alpha_\mathrm{L}$ from source2 are moderately worse than the results from source1 for all scenarios (single LISA or joint observation, unknowing or knowing the sky location) as two plots shown in the middle panel.
However, for the measurement on the $\alpha_\mathrm{x}$ and $\alpha_\mathrm{y}$, the results from source1 observations from the single LISA mission are still better than the results from source2, the joint observations could promote the source2 to a better constraint than source1 which could due to the TAIJI mission observe the source2 with a better response than source1. 
For the inclination $\iota=0.55 ~\mathrm{rad}$, the joint LISA-TAIJI observation could improve the parameter accuracy by more than 10 times compared to single LISA observation for $\alpha_\mathrm{L}$, $\alpha_\mathrm{x}$, and $\alpha_\mathrm{y}$; if the position of the source is known and excluded from FIM calculation, the uncertainties of parameters could further decrease, and this should be due to the degeneracy removed between the sky location and $\alpha_i$. 

\subsection{Results with varying ppE parameters $b$ and $\beta$}

\begin{figure*}[htb]
\includegraphics[width=0.46\textwidth]{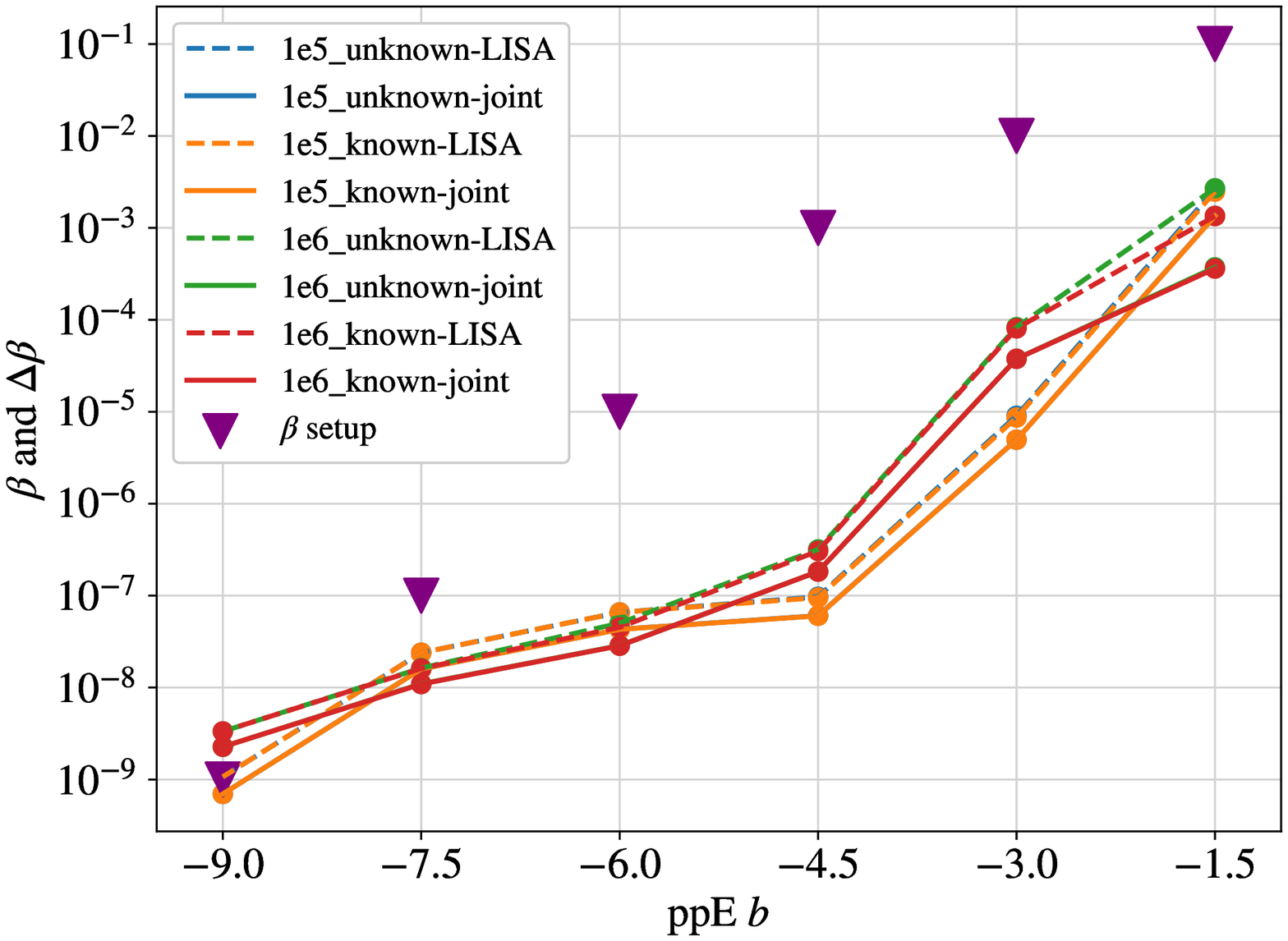} 
\includegraphics[width=0.46\textwidth]{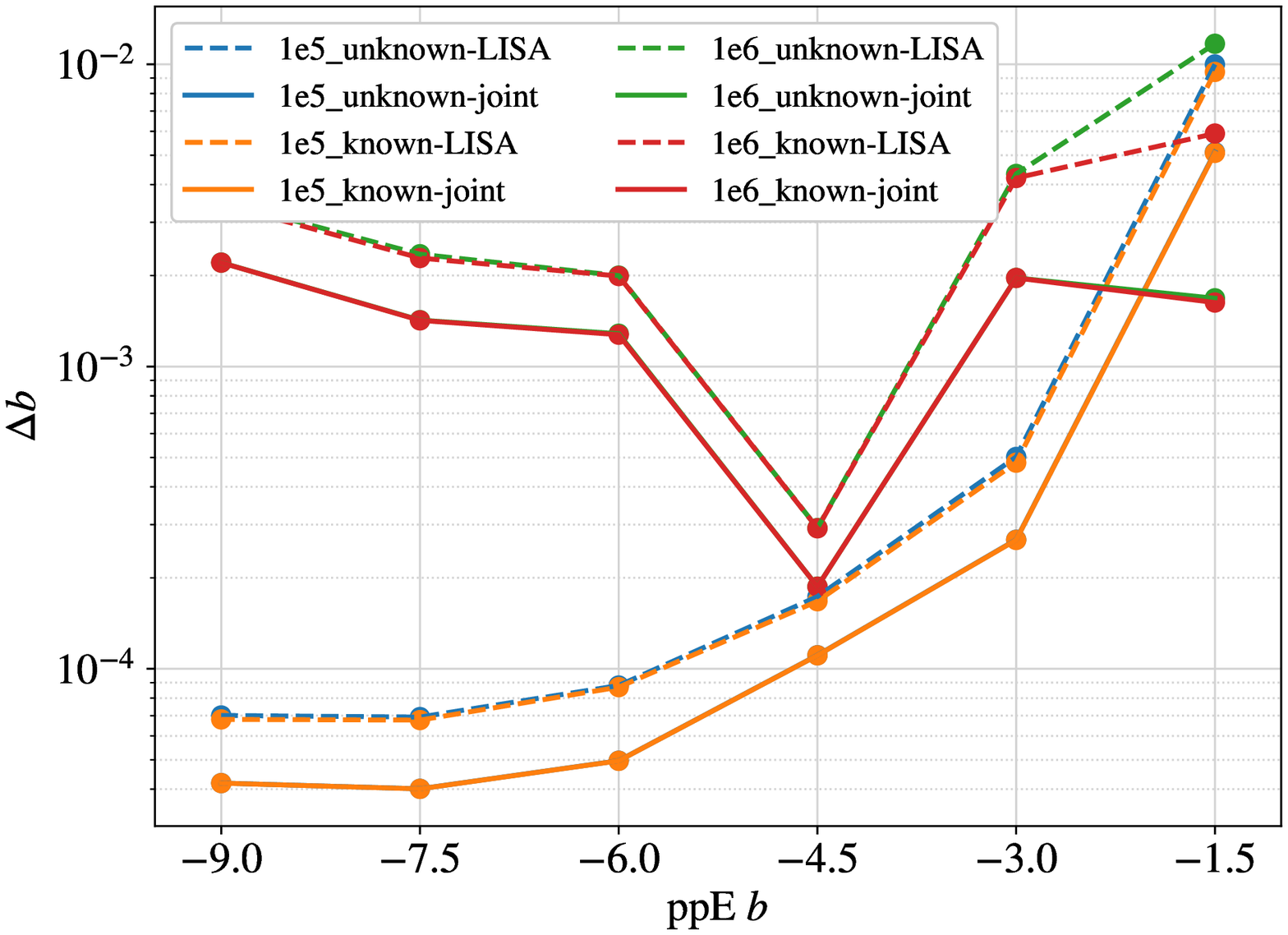} 
\includegraphics[width=0.46\textwidth]{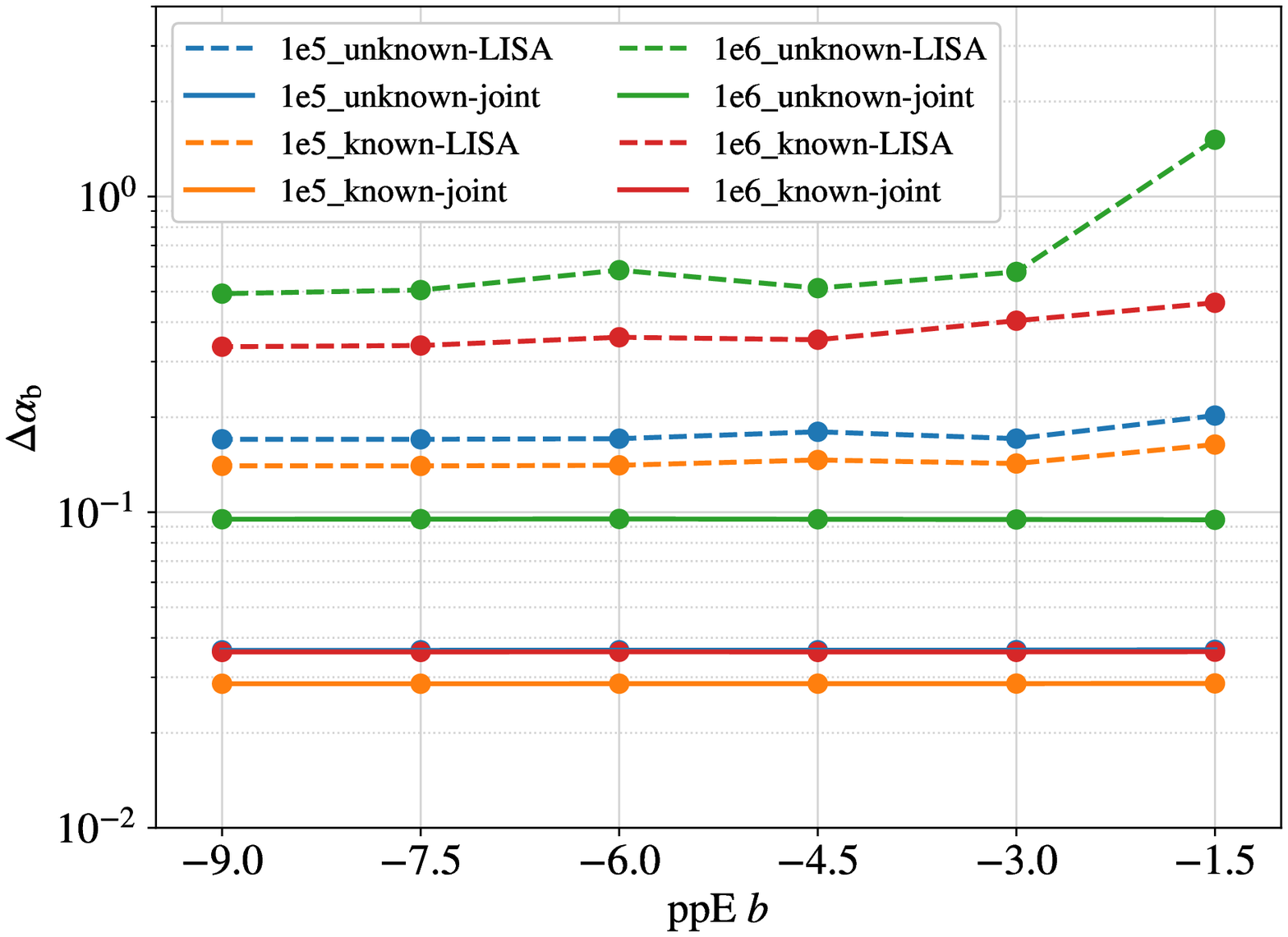} 
\includegraphics[width=0.46\textwidth]{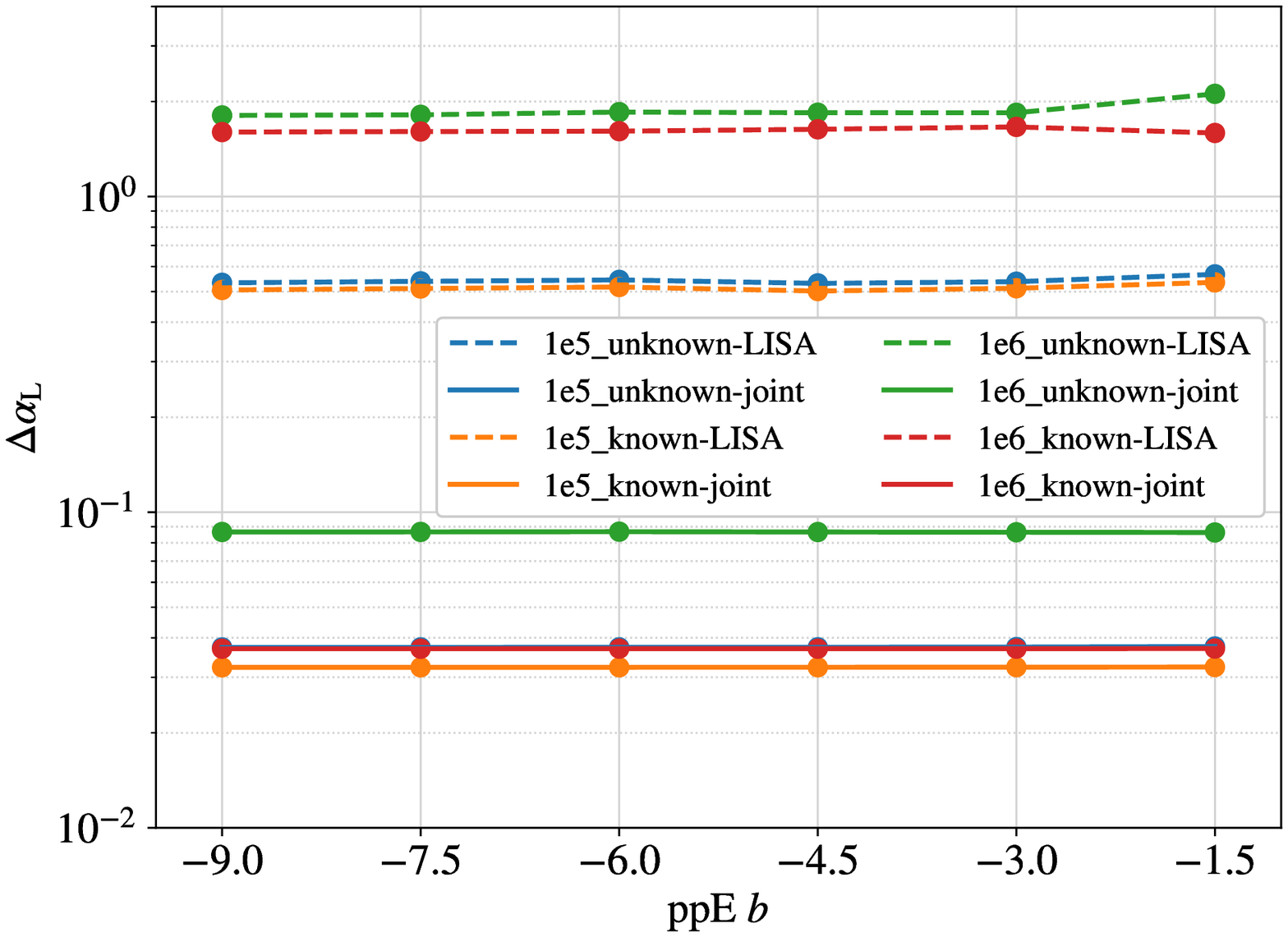} 
\includegraphics[width=0.46\textwidth]{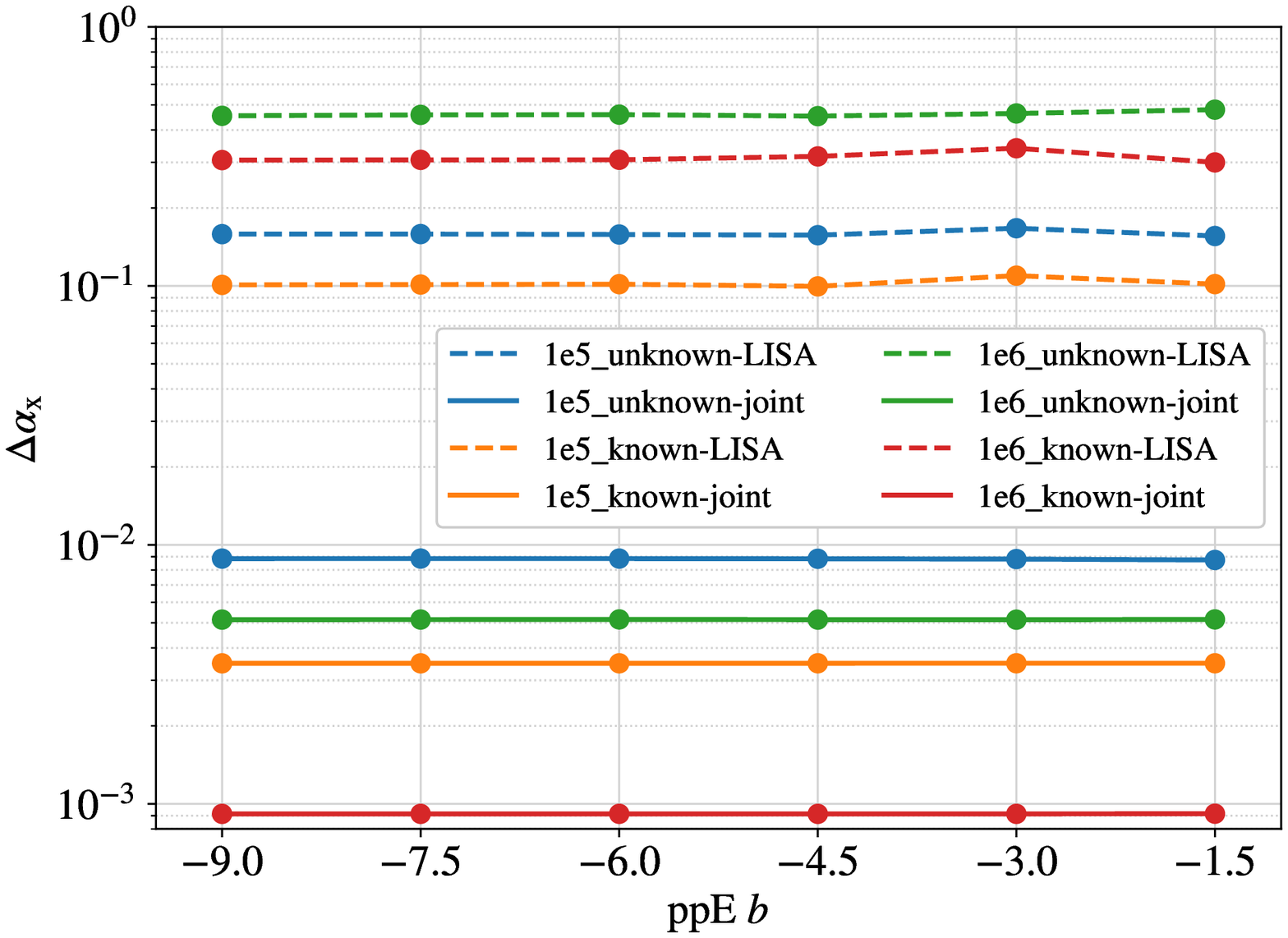} 
\includegraphics[width=0.46\textwidth]{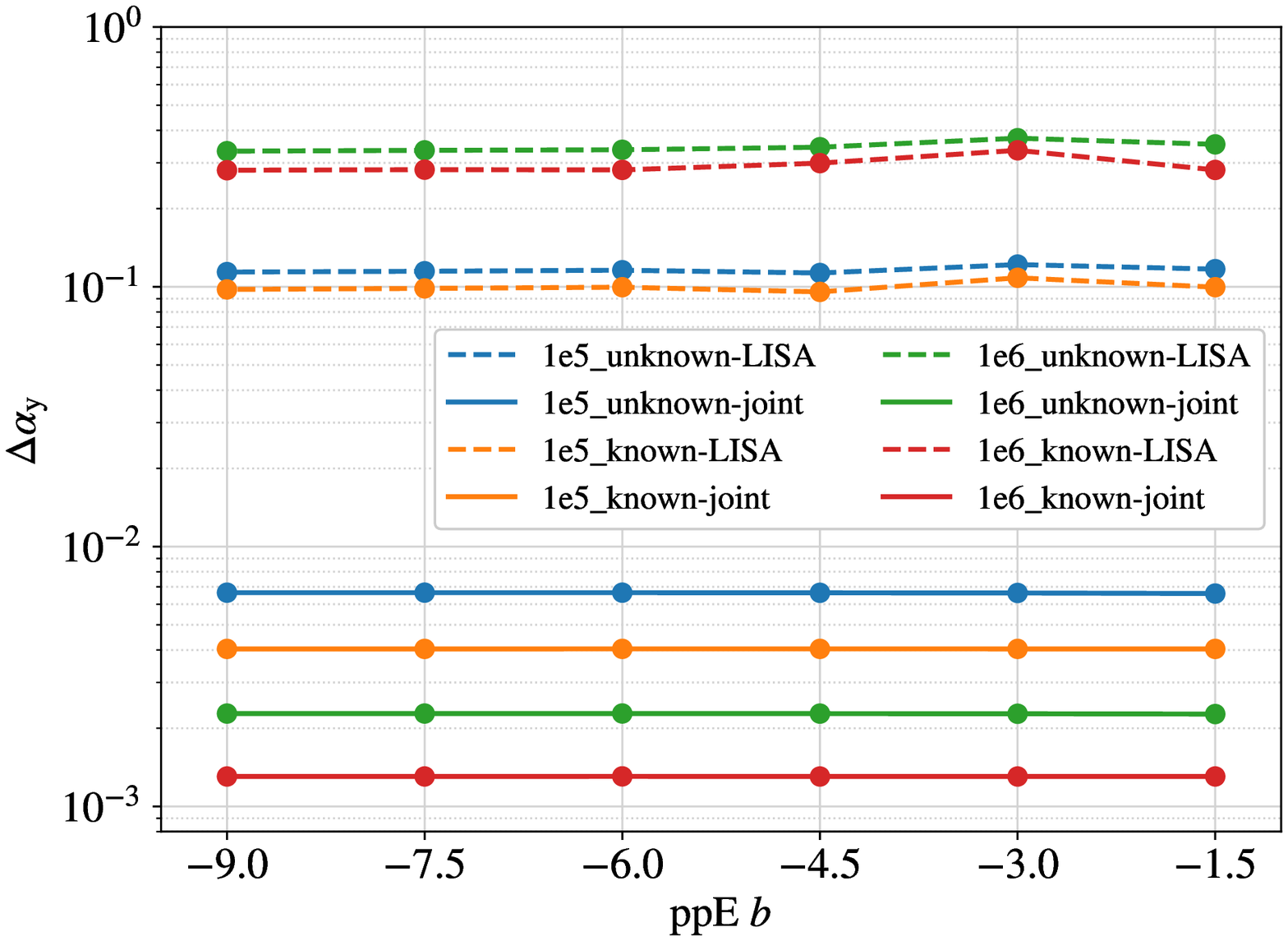} 
\caption{\label{fig:ppE_b_vs_beta_b}  The constraints of ppE parameters with different $\beta$ and $b$ from the LISA observation and LISA-TAIJI joint observations. The upper two plots show the results for $\beta$ and $b$, and the middle and lower panels are the results for alternative coefficients $\alpha_i$. The (purple) triangles are the $\beta$ setups at a given $b$ values which roughly referred from bounds in \citet{Cornish:2011ys}. The legend labels are same defined as in Fig. \ref{fig:iota_vs_ppE_paras}, the keyword with 1e5 or 1e6 indicates the respective source1 or source2, and a label with unknown/know shows if the position of source $(\lambda, \theta, D)$ is included/excluded in the FIM calculations. In upper two plots, the unknown and known curves are overlapped for each scenarios.
In the two plots in upper panel, the paired curves for $m_1$\_unkonwn-joint and $m_1$\_known-joint are overlapped.
}
\end{figure*}

In this subsection, we examine the impact of the joint LISA-TAIJI observation on the measurements of ppE parameters with different given $\beta$ and $b$ values.
As aforementioned, the inclination $\iota$ of a source tunes the amplitudes of each GW polarization and affects the SNR of the detection. 
Considering the tensor polarizations are dominant for the GW radiation from the coalescing compact binaries \cite{Abbott:2018lct,ligotestgr1,takeda2020arXiv201014538T}, we perform the investigations by choosing the fixed $\iota = 0.55 ~\mathrm{rad}$.
The four $\alpha_i$ coefficient on polarization amplitudes are set to be zero as fiducial value ($\alpha_\mathrm{b} = \alpha_\mathrm{L} = \alpha_\mathrm{x} = \alpha_\mathrm{x} = 0 $). Considering the parameter $\beta$ has been bound at a given $b$ from the PSR J0737-3039 \cite{Yunes:2010qb} and the LIGO and LISA simulation \cite{Cornish:2011ys}, our choices of $\beta$ at a given $b$ are shown by the purple triangles in the first plot of Fig. \ref{fig:ppE_b_vs_beta_b}. And the FIM is calculated subsequently by settling each pair $\beta$ and $b$. 

The constraints on the ppE parameters with different $\beta$ and $b$ are shown in Fig. \ref{fig:ppE_b_vs_beta_b}. The upper two plots show the results for $\beta$ and $b$ from the source1 ($m_1 = 10^5 M_{\odot}, q=1/3, z=2$) and source2 ($m_1 = 10^6 M_{\odot}, q=1/3, z=2$). As we can read from the upper left plots, for the presets of $\beta$ and $b$, the constraints on $\beta$ get better with the $b$ decreases for both two selected sources. The source1 shows relatively better determination on $\beta$ by around one order than the source2 for the ($\beta = 10^{-2}, b= -3$).
For other cases, no significant difference between their results. The joint observation of the LISA-TAIJI network could improve by a factor of $\sim$2 as shown more clearly in the previous subsection. 
For the measurement on parameter $b$, the source1 demonstrates more than 10 times better constraint than source2 for $b<-4.5$.  
And the joint observation also can improve by a factor of $\sim$2 on the parameter determination. The knowledge of the source position has a little improvement on the measurement since the sky location could be resolved from the loud signals, and their curves are overlapped with the unknown cases.
The measurements of $\alpha_i$ are shown in the middle and lower panels. Considering the $\alpha_i$ are relatively independent of the $\beta$ and $b$ selections, the constraint on the $\alpha_i$ are almost have no change with the $\beta$ and $b$ values.

We conclude that, for the ppE parameter $( \beta, b, \alpha_\mathrm{b}, \alpha_\mathrm{L}, \alpha_\mathrm{x}, \alpha_\mathrm{y})$ measurements from the selected sources, compared to the LISA single detector observation, the joint observation of the LISA and TAIJI network could improve for the $\beta$ and $b$ measurement by a factor of $\sim$2, the coefficients of alternative polarization modes $\alpha_i$ could be improved by more than $\sim$10 times. With knowing the position of the source, all the accuracy of polarization coefficients $\alpha_i$ could be further improved for some cases. For the significant promotions on the measurement of $\alpha_i$, we ascribe the sky coverage compensation for LISA and TAIJI missions as inferred in Fig. \ref{fig:molllweide_polarization}. For our selected locations of the GW sources, the LISA could efficiently observe the GW tensor polarization modes and insensitively detect the other polarization modes. As a merit of the joint network, the TAIJI mission could response to the alternative polarizations with a better antenna pattern in this case and observe these polarizations with a higher sensitivity. A caveat is that the different choices about the source location and/or merger time may yield a different constraints on the ppE parameters.

\section{Conclusions} \label{sec:conclusions}

In this work, we explore the detectability of the LISA-TAIJI network to the alternative polarization modes compared to that of the single LISA mission. The ppE formulation is employed to specify the parameters $(\beta, b, \alpha_\mathrm{b}, \alpha_\mathrm{L}, \alpha_\mathrm{x}, \alpha_\mathrm{y})$ to be determined. To perform the investigations, two sources are selected which are source1 ($m_1=10^5 \ M_\odot, q =1/3$) at redshift $z=2$, and source2 ($m_1 = 10^6 \ M_\odot, q=1/3$) at the same distance. By using the Fisher matrix algorithm,
for the last one year to coalescence, the ppE parameters are generally better measured from the source1 observation than the source2.  

The joint LISA-TAIJI network could improve the measurement of $\beta$ and $b$ by a factor of $\sim$2 compared to the single LISA mission.
The joint observations show the significant improvement for the uncertainty of the alternative polarization modes coefficients $\alpha_i$, and the joint network could reduce the uncertainty of the $\alpha_i$ by a factor of $\gtrsim10$ compared to LISA except $\gtrsim 4$ for $\alpha_\mathrm{b}$. In an optimistic scenario, if the location of the source is determined by the multi-messenger observation, the joint LISA-TAIJI observation could further improve the measurement of four coefficients of alternative polarization modes $\alpha_i$, which should be an outcome of removing the degeneracies between the source distance and coefficients. 

The current study employs the Fisher information matrix algorithm to determine the uncertainties of the ppE parameters with a single event, and only the approximate limits are achieved from this investigation. The Bayesian approaches have been proposed by \citet{DelPozzo:2011pg} and \citet{Cornish:2011ys} to test the alternative gravitational theories. And more rigorous bounds could be obtained by applying the Bayesian algorithm to the LISA-TAIJI joint observation. 
We plan to perform these analyses in future studies.

\begin{acknowledgments}
We thank Prof. Wei-Tou Ni for helpful discussions and comments, and we also thank the anonymous referee for the constructive comments in helping us improve the manuscript. This work was supported by NSFC Nos. 12003059 and 11773059, Key Research Program of Frontier Sciences, Chinese Academy of Science, No. QYZDB-SSW-SYS016 and the Strategic Priority Research Program of the Chinese Academy of Sciences under grant Nos. XDA1502070102, XDA15020700 and XDB21010100. and by the National Key Research and Development Program of China under Grant Nos. 2016YFA0302002 and 2017YFC0601602. This work made use of the High Performance Computing Resource in the Core Facility for Advanced Research Computing at Shanghai Astronomical Observatory.
\end{acknowledgments}

\appendix
\section{Appendix} \label{sec:appendix}

\begin{figure*}[htb]
\includegraphics[width=0.16\textwidth]{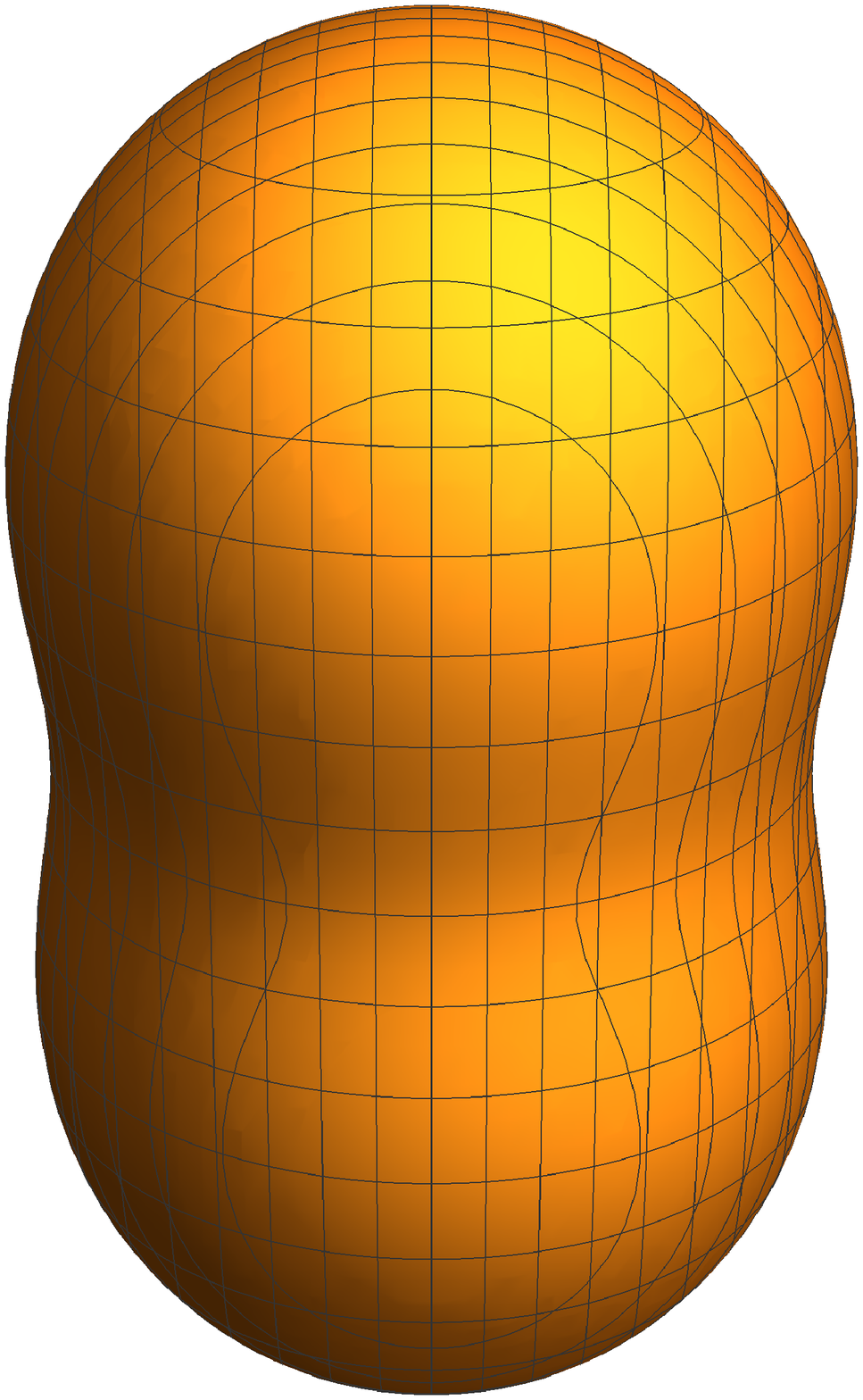} 
\includegraphics[width=0.16\textwidth]{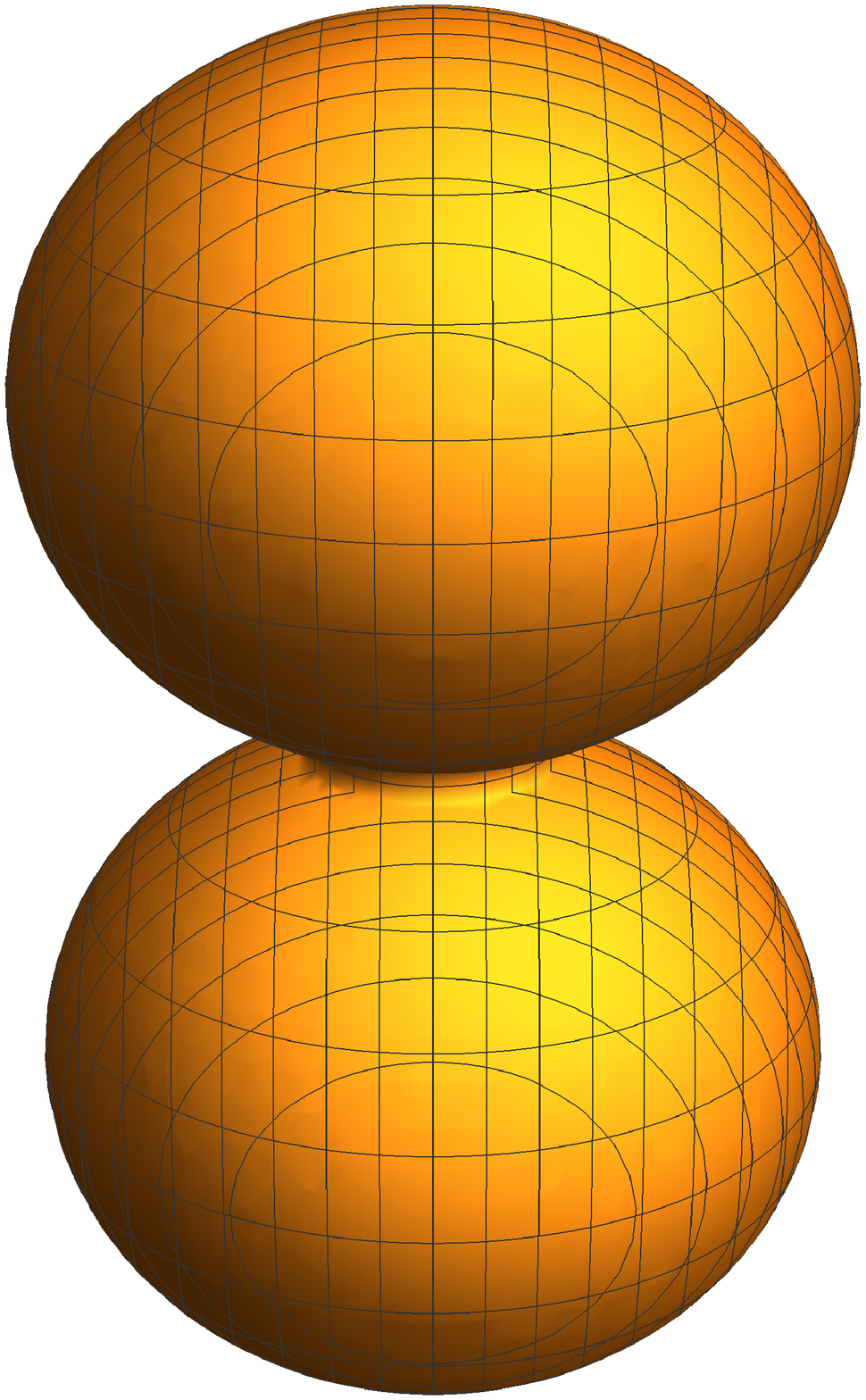}
\includegraphics[width=0.2\textwidth]{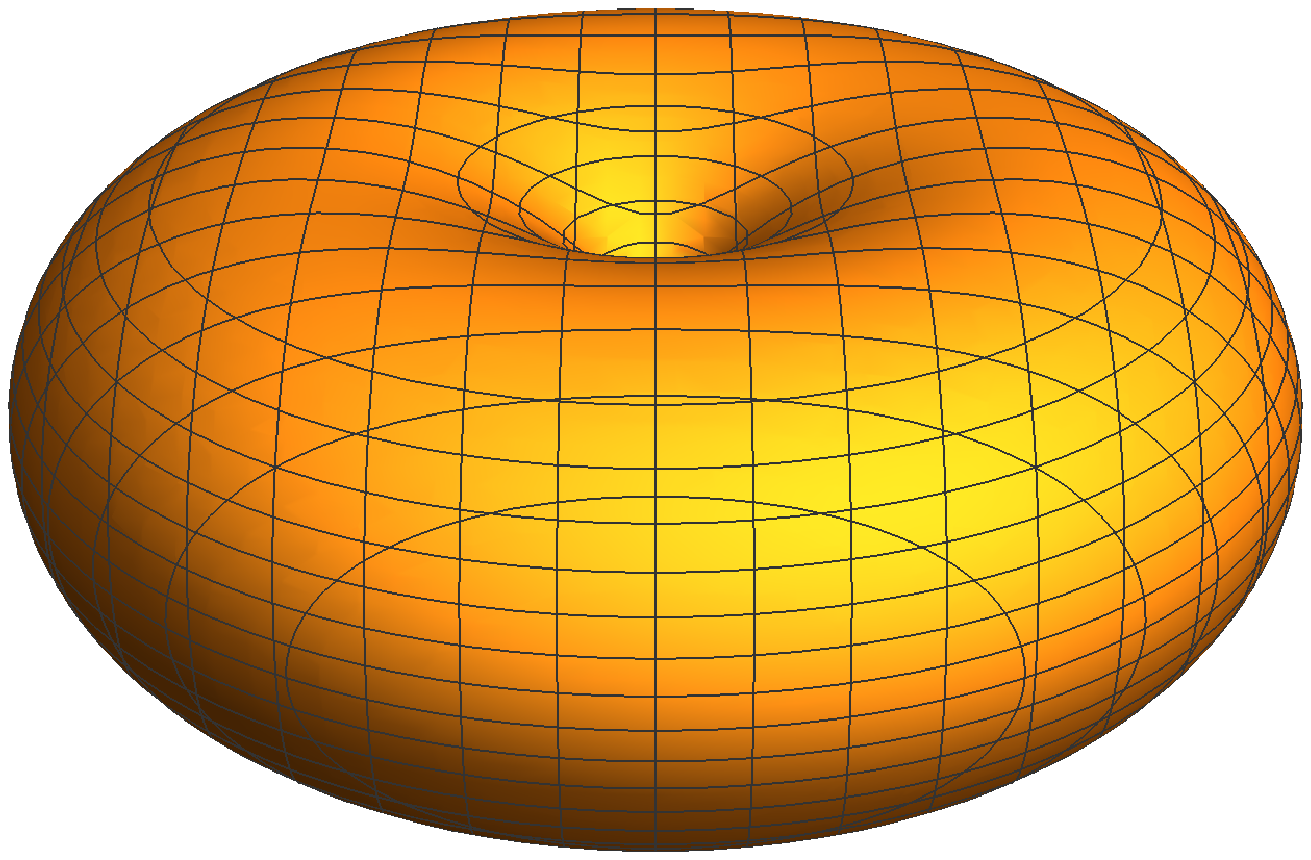} 
\includegraphics[width=0.2\textwidth]{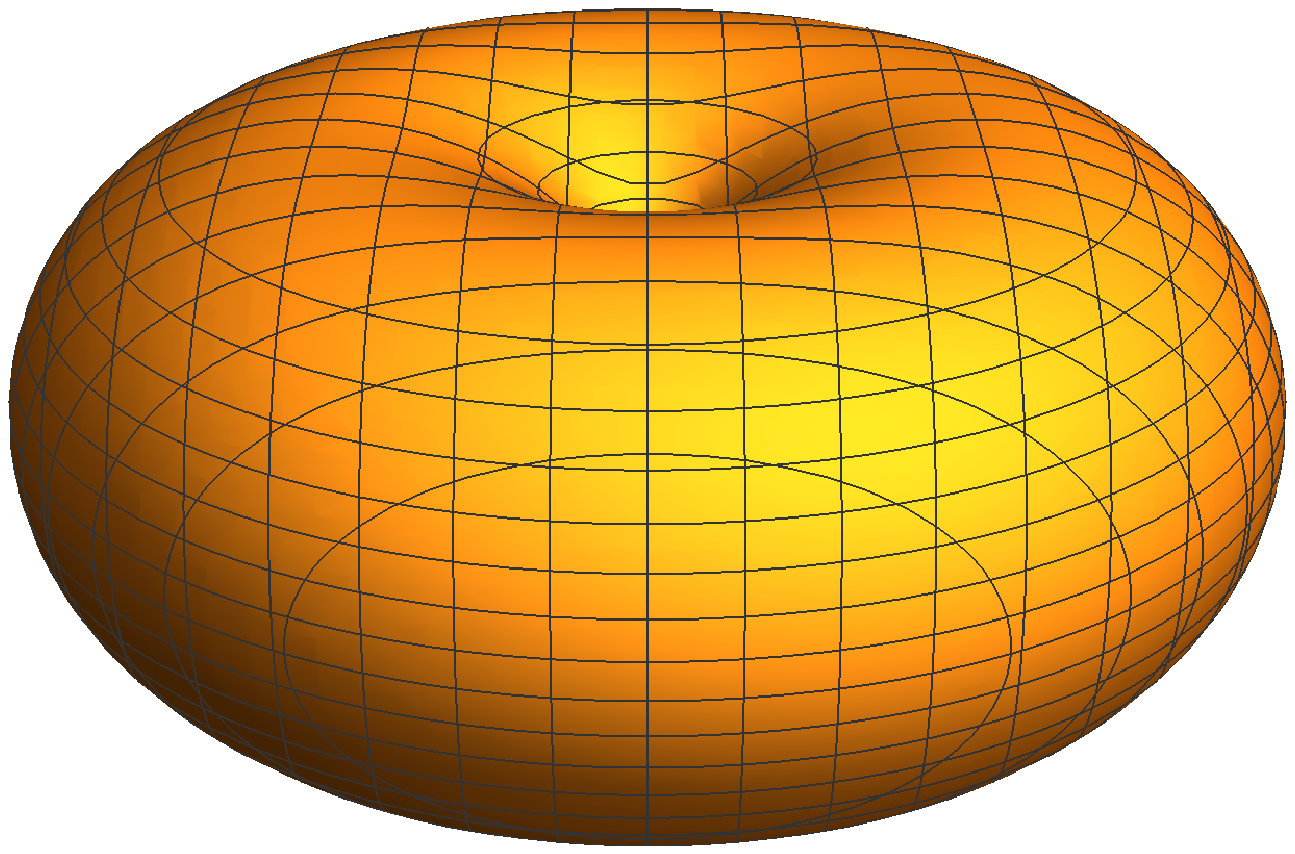}
\includegraphics[width=0.2\textwidth]{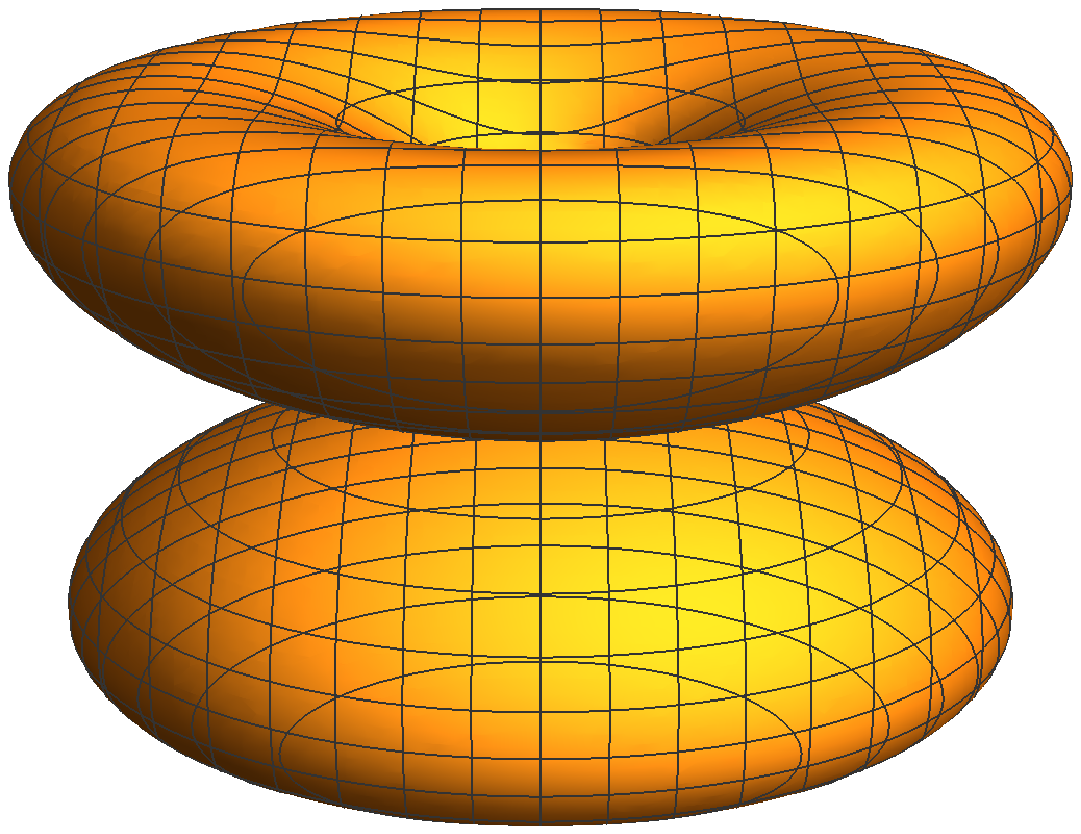}
\caption{\label{fig:antenna_polarization} The antenna pattern of joint LISA A+E+T channels for polarization modes tensor $+$ (first plot), tensor $\times$ (second plot), scalar b/L (third plot), vector x (forth plot) and vector y (last plot) at 10 mHz in the detector frame.}
\end{figure*}

\nocite{*}
\bibliography{apsref}

\end{document}